\documentclass[aps,prd,twocolumn,superscriptaddress,showpacs,floatfix,nofootinbib]{revtex4-1}
\usepackage{graphicx}

\usepackage{subfigure}
\usepackage{xspace}
\usepackage{dcolumn}
\usepackage{multirow}
\usepackage{longtable}
\usepackage{amsmath}
\usepackage{mathtools}
\usepackage{xcolor}
\usepackage{lineno}
\usepackage{grffile}
\usepackage{hyperref}

\begin{document}

\title{Search for proton decay via $p\rightarrow \mu^+K^0$ in 0.37 megaton-years exposure of Super-Kamiokande}

\newcommand{\AFFicrr}{\affiliation{Kamioka Observatory, Institute for Cosmic Ray Research, University of Tokyo, Kamioka, Gifu 506-1205, Japan}}
\newcommand{\AFFkashiwa}{\affiliation{Research Center for Cosmic Neutrinos, Institute for Cosmic Ray Research, University of Tokyo, Kashiwa, Chiba 277-8582, Japan}}
\newcommand{\AFFicrronly}{\affiliation{Institute for Cosmic Ray Research, University of Tokyo, Kashiwa, Chiba 277-8582, Japan}}
\newcommand{\AFFipmu}{\affiliation{Kavli Institute for the Physics and
Mathematics of the Universe (WPI), The University of Tokyo Institutes for Advanced Study,
University of Tokyo, Kashiwa, Chiba 277-8583, Japan }}
\newcommand{\AFFmad}{\affiliation{Department of Theoretical Physics, University Autonoma Madrid, 28049 Madrid, Spain}}
\newcommand{\AFFubc}{\affiliation{Department of Physics and Astronomy, University of British Columbia, Vancouver, BC, V6T1Z4, Canada}}
\newcommand{\AFFbu}{\affiliation{Department of Physics, Boston University, Boston, MA 02215, USA}}
\newcommand{\AFFuci}{\affiliation{Department of Physics and Astronomy, University of California, Irvine, Irvine, CA 92697-4575, USA }}
\newcommand{\AFFcsu}{\affiliation{Department of Physics, California State University, Dominguez Hills, Carson, CA 90747, USA}}
\newcommand{\AFFcnm}{\affiliation{Institute for Universe and Elementary Particles, Chonnam National University, Gwangju 61186, Korea}}
\newcommand{\AFFduke}{\affiliation{Department of Physics, Duke University, Durham NC 27708, USA}}
\newcommand{\AFFfukuoka}{\affiliation{Junior College, Fukuoka Institute of Technology, Fukuoka, Fukuoka 811-0295, Japan}}
\newcommand{\AFFgifu}{\affiliation{Department of Physics, Gifu University, Gifu, Gifu 501-1193, Japan}}
\newcommand{\AFFgist}{\affiliation{GIST College, Gwangju Institute of Science and Technology, Gwangju 500-712, Korea}}
\newcommand{\AFFuh}{\affiliation{Department of Physics and Astronomy, University of Hawaii, Honolulu, HI 96822, USA}}
\newcommand{\AFFicl}{\affiliation{Department of Physics, Imperial College London , London, SW7 2AZ, United Kingdom }}
\newcommand{\AFFkek}{\affiliation{High Energy Accelerator Research Organization (KEK), Tsukuba, Ibaraki 305-0801, Japan }}
\newcommand{\AFFkobe}{\affiliation{Department of Physics, Kobe University, Kobe, Hyogo 657-8501, Japan}}
\newcommand{\AFFkyoto}{\affiliation{Department of Physics, Kyoto University, Kyoto, Kyoto 606-8502, Japan}}
\newcommand{\AFFliv}{\affiliation{Department of Physics, University of Liverpool, Liverpool, L69 7ZE, United Kingdom}}
\newcommand{\AFFmiyagi}{\affiliation{Department of Physics, Miyagi University of Education, Sendai, Miyagi 980-0845, Japan}}
\newcommand{\AFFnagoya}{\affiliation{Institute for Space-Earth Environmental Research, Nagoya University, Nagoya, Aichi 464-8602, Japan}}
\newcommand{\AFFkmi}{\affiliation{Kobayashi-Maskawa Institute for the Origin of Particles and the Universe, Nagoya University, Nagoya, Aichi 464-8602, Japan}}
\newcommand{\AFFpol}{\affiliation{National Centre For Nuclear Research, 02-093 Warsaw, Poland}}
\newcommand{\AFFsuny}{\affiliation{Department of Physics and Astronomy, State University of New York at Stony Brook, NY 11794-3800, USA}}
\newcommand{\AFFokayama}{\affiliation{Department of Physics, Okayama University, Okayama, Okayama 700-8530, Japan }}
\newcommand{\AFFosaka}{\affiliation{Department of Physics, Osaka University, Toyonaka, Osaka 560-0043, Japan}}
\newcommand{\AFFox}{\affiliation{Department of Physics, Oxford University, Oxford, OX1 3PU, United Kingdom}}
\newcommand{\AFFqmul}{\affiliation{School of Physics and Astronomy, Queen Mary University of London, London, E1 4NS, United Kingdom}}
\newcommand{\AFFregina}{\affiliation{Department of Physics, University of Regina, 3737 Wascana Parkway, Regina, SK, S4SOA2, Canada}}
\newcommand{\AFFseoul}{\affiliation{Department of Physics, Seoul National University, Seoul 151-742, Korea}}
\newcommand{\AFFsheff}{\affiliation{Department of Physics and Astronomy, University of Sheffield, S3 7RH, Sheffield, United Kingdom}}
\newcommand{\AFFshizuokasc}{\affiliation{Department of Informatics in
Social Welfare, Shizuoka University of Welfare, Yaizu, Shizuoka, 425-8611, Japan}}
\newcommand{\AFFstfc}{\affiliation{STFC, Rutherford Appleton Laboratory, Harwell Oxford, and Daresbury Laboratory, Warrington, OX11 0QX, United Kingdom}}
\newcommand{\AFFskk}{\affiliation{Department of Physics, Sungkyunkwan University, Suwon 440-746, Korea}}
\newcommand{\AFFtokyo}{\affiliation{The University of Tokyo, Bunkyo, Tokyo 113-0033, Japan }} 
\newcommand{\AFFtodai}{\affiliation{Department of Physics, University of Tokyo, Bunkyo, Tokyo 113-0033, Japan }}
\newcommand{\AFFtit}{\affiliation{Department of Physics,Tokyo Institute of Technology, Meguro, Tokyo 152-8551, Japan }}
\newcommand{\AFFtus}{\affiliation{Department of Physics, Faculty of Science and Technology, Tokyo University of Science, Noda, Chiba 278-8510, Japan }}
\newcommand{\AFFtoronto}{\affiliation{Department of Physics, University of Toronto, ON, M5S 1A7, Canada }}
\newcommand{\AFFtriumf}{\affiliation{TRIUMF, 4004 Wesbrook Mall, Vancouver, BC, V6T2A3, Canada }}
\newcommand{\AFFtokai}{\affiliation{Department of Physics, Tokai University, Hiratsuka, Kanagawa 259-1292, Japan}}
\newcommand{\AFFtsinghua}{\affiliation{Department of Engineering Physics, Tsinghua University, Beijing, 100084, China}}
\newcommand{\AFFynu}{\affiliation{Department of Physics, Yokohama National University, Yokohama, Kanagawa, 240-8501, Japan}}
\newcommand{\AFFllr}{\affiliation{Ecole Polytechnique, IN2P3-CNRS, Laboratoire Leprince-Ringuet, F-91120 Palaiseau, France }}
\newcommand{\AFFbari}{\affiliation{ Dipartimento Interuniversitario di Fisica, INFN Sezione di Bari and Universit\`a e Politecnico di Bari, I-70125, Bari, Italy}}
\newcommand{\AFFnapoli}{\affiliation{Dipartimento di Fisica, INFN Sezione di Napoli and Universit\`a di Napoli, I-80126, Napoli, Italy}}
\newcommand{\AFFroma}{\affiliation{INFN Sezione di Roma and Universit\`a di Roma ``La Sapienza'', I-00185, Roma, Italy}}
\newcommand{\AFFpadova}{\affiliation{Dipartimento di Fisica, INFN Sezione di Padova and Universit\`a di Padova, I-35131, Padova, Italy}}
\newcommand{\AFFkeio}{\affiliation{Department of Physics, Keio University, Yokohama, Kanagawa, 223-8522, Japan}}
\newcommand{\AFFwinnipeg}{\affiliation{Department of Physics, University of Winnipeg, MB R3J 3L8, Canada }}
\newcommand{\AFFkcl}{\affiliation{Department of Physics, King's College London, London, WC2R 2LS, UK }}
\newcommand{\AFFwarwick}{\affiliation{Department of Physics, University of Warwick, Coventry, CV4 7AL, UK }}
\newcommand{\AFFral}{\affiliation{Rutherford Appleton Laboratory, Harwell, Oxford, OX11 0QX, UK }}
\newcommand{\AFFwu}{\affiliation{Faculty of Physics, University of Warsaw, Warsaw, 02-093, Poland }}
\newcommand{\AFFbcit}{\affiliation{Department of Physics, British Columbia Institute of Technology, Burnaby, BC, V5G 3H2, Canada }}
\newcommand{\AFFtohoku}{\affiliation{Department of Physics, Faculty of Science, Tohoku University, Sendai, Miyagi, 980-8578, Japan }}
\newcommand{\AFFicise}{\affiliation{Institute For Interdisciplinary Research in Science and Education, ICISE, Quy Nhon, 55121, Vietnam }}
\newcommand{\AFFilance}{\affiliation{ILANCE, CNRS - University of Tokyo International Research Laboratory, Kashiwa, Chiba 277-8582, Japan}}
\newcommand{\AFFibs}{\affiliation{Institute for Basic Science (IBS), Daejeon, 34126, Korea}}

\AFFicrr
\AFFkashiwa
\AFFicrronly
\AFFmad
\AFFbu
\AFFbcit
\AFFuci
\AFFcsu
\AFFcnm
\AFFduke
\AFFllr
\AFFfukuoka
\AFFgifu
\AFFgist
\AFFuh
\AFFibs
\AFFicise
\AFFicl
\AFFbari
\AFFnapoli
\AFFpadova
\AFFroma
\AFFilance
\AFFkeio
\AFFkek
\AFFkcl
\AFFkobe
\AFFkyoto
\AFFliv
\AFFmiyagi
\AFFnagoya
\AFFkmi
\AFFpol
\AFFsuny
\AFFokayama
\AFFox
\AFFral
\AFFseoul
\AFFsheff
\AFFshizuokasc
\AFFstfc
\AFFskk
\AFFtohoku
\AFFtokai
\AFFtokyo 
\AFFtodai
\AFFipmu
\AFFtit
\AFFtus
\AFFtoronto
\AFFtriumf
\AFFtsinghua
\AFFwu
\AFFwarwick
\AFFwinnipeg
\AFFynu

\author{R.~Matsumoto}
\AFFtus

\author{K.~Abe}
\AFFicrr
\AFFipmu
\author{Y.~Hayato}
\AFFicrr
\AFFipmu
\author{K.~Hiraide}
\AFFicrr
\AFFipmu
\author{K.~Ieki}
\author{M.~Ikeda}
\AFFicrr
\AFFipmu
\author{J.~Kameda}
\AFFicrr
\AFFipmu
\author{Y.~Kanemura}
\author{R.~Kaneshima}
\author{Y.~Kashiwagi}
\AFFicrr
% ...


\collaboration{The Super-Kamiokande Collaboration}
\noaffiliation

\date{\today}

\begin{abstract}
    We searched for proton decay via $p\to\mu^+K^0$ in 0.37\,Mton$\cdot$years of data collected between 1996 and 2018 from the Super-Kamiokande water Cherenkov experiment. 
    The selection criteria were defined separately for $K^0_S$ and $K^0_L$ channels.
    No significant event excess has been observed.
    As a result of this analysis, which extends the previous search by an additional 0.2\,Mton$\cdot$years of exposure and uses an improved event reconstruction, we set a lower limit of $3.6\times10^{33}$ years on the proton lifetime.
\end{abstract}
\maketitle

\section{Introduction}
A Grand Unified Theory (GUT) is an extension of the Standard Model (SM) of particle physics that unifies leptons, baryons and gauge bosons \cite{GUT}.
GUTs often lead to proton decay through interactions in which baryon number is not conserved.
The dominant proton decay channel depends on the assumptions of the framework and parameters.
Based on experimental searches for $p\to e^+\pi^0$ \cite{IMB_epi0}\cite{Kamiokande_epi0}\cite{SK_epi0_3}, the simplest minimal SU(5) GUT model \cite{minimal_SU5} has already been rejected.
Supersymmetric (SUSY) GUT models often predict $p\to\bar{\nu}K^+$ and $p\to\mu^+K^0$ as the dominant channel due to the contribution of the color triplet Higgs.
The partial lifetime of the $p\to\mu^+K^0$ channel can be comparable to $p\to\bar{\nu}K^+$ or even shorter.
Some models \cite{pmuK0_theory} predict the partial lifetime for $p \rightarrow \mu^+ K^0$ to be in a range that slightly exceeds the current experimental limit of $1.6 \times 10^{33}$ years \cite{Regis}.  These models generally prefer shorter lifetimes accessible to our current search.

In this paper, we focus on a search for the $p\to\mu^+K^0$ mode at the Super-Kamiokande (SK) detector.
Due to the large fiducial mass and ability to identify the particle type and measure its momentum, SK is well suited to search for this decay channel.
This two-body decay generates a muon and a neutral kaon with monochromatic momentum of 326.5\,MeV$/c$.
The $K^0$ is a composite state of $K^0_S$ and $K^0_L$.
$K^0_S$ decays into $\pi^+\pi^-$ (69.2\%) and $2\pi^0$ (30.7\%) with a lifetime of 90\,ps while $K^0_L$ decays into $\pi^\pm e^\mp \nu$ (40.6\%), $\pi^\pm \mu^\mp \nu$ (27.0\%), $3\pi^0$ (19.5\%) and $\pi^+\pi^-\pi^0$ (12.5\%) with the lifetime of 51\,ns.

Proton decay searches for the $p \rightarrow \mu^+ K^0$ mode have been performed in SK \cite{Kobayashi_muK0}\cite{Regis}.
In the previous search by Super-Kamiokande, the full data of the SK-I, SK-II, and SK-III phases (0.17 Mton$\cdot$years of exposure from 1996 to 2008) were analyzed and no significant signal of $p \rightarrow \mu^+ K^0$ was observed \cite{Regis}. A lower limit on the partial proton lifetime $p \rightarrow \mu^+ K^0$ of $1.6 \times 10^{33}$ years was set at the 90\% confidence level (C.L.). In this paper, the search was extended with the full data set of the SK-IV phase, adding 0.20 Mton$\cdot$years of exposure from 2008 to 2018. For SK-IV, upgraded electronics and data acquisition systems \cite{SK4_electronics} allow for the detection of neutron capture on hydrogen. In addition to neutron tagging, the SK-IV data were analyzed with other improvements in event reconstruction and newly optimized selection criteria. In this paper, we will statistically combine the result of the SK-IV search with the previously published result, taking common systematic uncertainties into account.

In this paper, the Super-Kamiokande detector and the simulation are described in Section \ref{sec:SK} and \ref{sec:simulation}, respectively.
Then, the improved event reconstruction and the neutron tagging are explained in Section \ref{sec:reconstruction}.
The event selection and the results are described in Section \ref{sec:selection} and \ref{sec:result}, respectively.
The lifetime limit is calculated in Section \ref{sec:lifetime} with the systematic uncertainties described in Section \ref{sec:syst}.
The conclusion is written in Section \ref{sec:conclusion}.

\section{Super-Kamiokande}
\label{sec:SK}
Super-Kamiokande is a large water Cherenkov detector located 1\,km underground (2,700 m.w.e.) in the Gifu Prefecture, Japan.
The detector is a cylindrical water tank with 39.4\,m diameter and 41.4\,m height filled with 50\,kton of ultra-pure water.
The water tank is divided into two concentric volumes that are optically separated. The inner detector (ID) is a cylindrical volume with 33.8\,m diameter and 36.2\,m height. It contains 32\,kton of water.
There are 11,129 inward-facing photomultiplier tubes (PMTs) with 50\,cm diameter on the wall to detect the Cherenkov light from charged particles. 
The photocathode coverage is about 40\%.
The ID is surrounded by a 2\,m thickness layer called the outer detector (OD).
The OD is composed of 1,885 outward-facing PMTs with 20-cm diameter and wavelength shifting plates.
The walls of the OD are covered with a reflective sheet made of a material called Tyvek to improve the light collection. 
The PMTs in the OD are mainly used to identify incoming cosmic ray muons and particles exiting the ID.

The data are divided into the periods from 1996 to 2001 (SK-I), from 2002 to 2005 (SK-II), from 2006 to 2008 (SK-III) and from 2008 to 2018 (SK-IV).
Photocathode coverage was 40\% except for the SK-II period with 20\% coverage.
Starting from SK-IV, new electronics were installed \cite{SK4_electronics} with which neutron tagging was realized even with pure water, as described in Section \ref{sec:reconstruction}.
In this study, data collected in SK-IV (3244.39\,live-days) were newly analyzed and the result was combined with the previous result using SK-I to SK-III (2805.9\,live-days in total) \cite{Regis}.
The detailed descriptions and calibrations of the detector can be found in \cite{SK_calib}.

\section{Simulation}
\label{sec:simulation}
Signal efficiencies and expected background rates were estimated by Monte Carlo (MC) simulations.
Signal efficiencies are estimated based on a MC sample with $10^5$ events.
Atmospheric neutrino MC samples based on 500 years of simulated data were used to estimate the expected background rate.
Propagation of particles in water is simulated by a GEANT3-based package \cite{Geant3}.
The generation and propagation of Cherenkov light and the response of PMT and electronics are simulated by a custom code.
For the scattering of hadrons in water and nucleus, pions are simulated by the NEUT program \cite{NEUT} and kaons are simulated by the custom code as explained later.

\subsection{Proton decay}
Proton decay events were generated from nuclei of oxygen and hydrogen in water within the ID.
Protons in oxygen interact with other nucleons and are called bound protons.
These bound protons experience Fermi motion.
Due to Fermi motion, momentum distributions of $\mu^+$ and $K^0$ from the decays of bound protons are different than those from free protons.
In contrast, protons in hydrogen decay with zero momentum.
In this case, the momenta of $\mu^+$ and $K^0$ are exactly the same and their directions are back-to-back.

%
The Fermi momentum distribution in the simulation is extracted using the measurement of electron-${}^{12}\mathrm{C}$ scattering \cite{Nakamura_12Cscat}.
The effective proton mass $m_p'$ is calculated by subtracting the binding energy $E_b$ from the proton rest mass $m_p$, where $E_b$ is simulated for each nuclear state as a Gaussian random variable with a mean and a standard deviation of 39.0\,MeV and 10.2\,MeV for the $s$-state and 15.5\,MeV and 3.82\,MeV for the $p$-state, respectively.
The ratio of protons in the $s$-state and $p$-state is taken to be 1:3 based on the nuclear shell model \cite{Mayer_nuclear_state}.
%
The effective mass of the bound proton is smaller than for free protons due to interactions with the surrounding nucleons. 
This correlated decay has an estimated probability 10\% of each proton decay \cite{Yamazaki_correlated_decay}.
%
If a kaon is generated in the nucleus by proton decay, it can interact with the surrounding nucleons in the nucleus and the water.
The kaon interaction model was updated from that used in the previously published SK-I to SK-III search as explained below.

In the nucleus, neutral kaon scattering is simulated assuming an eigenstate of $K^0$.
The scattering probability of $K^0$ is calculated from the mean free path for every 0.2\,fm step in the nucleus assuming the nucleon density of oxygen \cite{Oxygen_density}.
Since $K^+$ and $K^0$ form an isospin doublet, the scattering amplitude of $K^0$ and a nucleon can be evaluated from the experimental results of $K^+$ scattering \cite{Hyslop_Kaon_CS}.
Elastic scattering and charge exchange are simulated in the nucleus as they are relevant for $K^0N$ interactions at the proton decay energy scale.
If the momentum of the scattered nucleon is below the Fermi surface momentum, scattering is suppressed by the Pauli exclusion principle.
The cross section for ${}^{16}\mathrm{O}$ is scaled to be proportional to the number of nucleons, referring to the experimental results \cite{Weiss}.

\begin{figure}[htbp]
    \centering
    \begin{tabular}{c}
        \begin{minipage}{1.0\hsize}
            \centering
            \includegraphics[keepaspectratio, width=1.0\hsize]{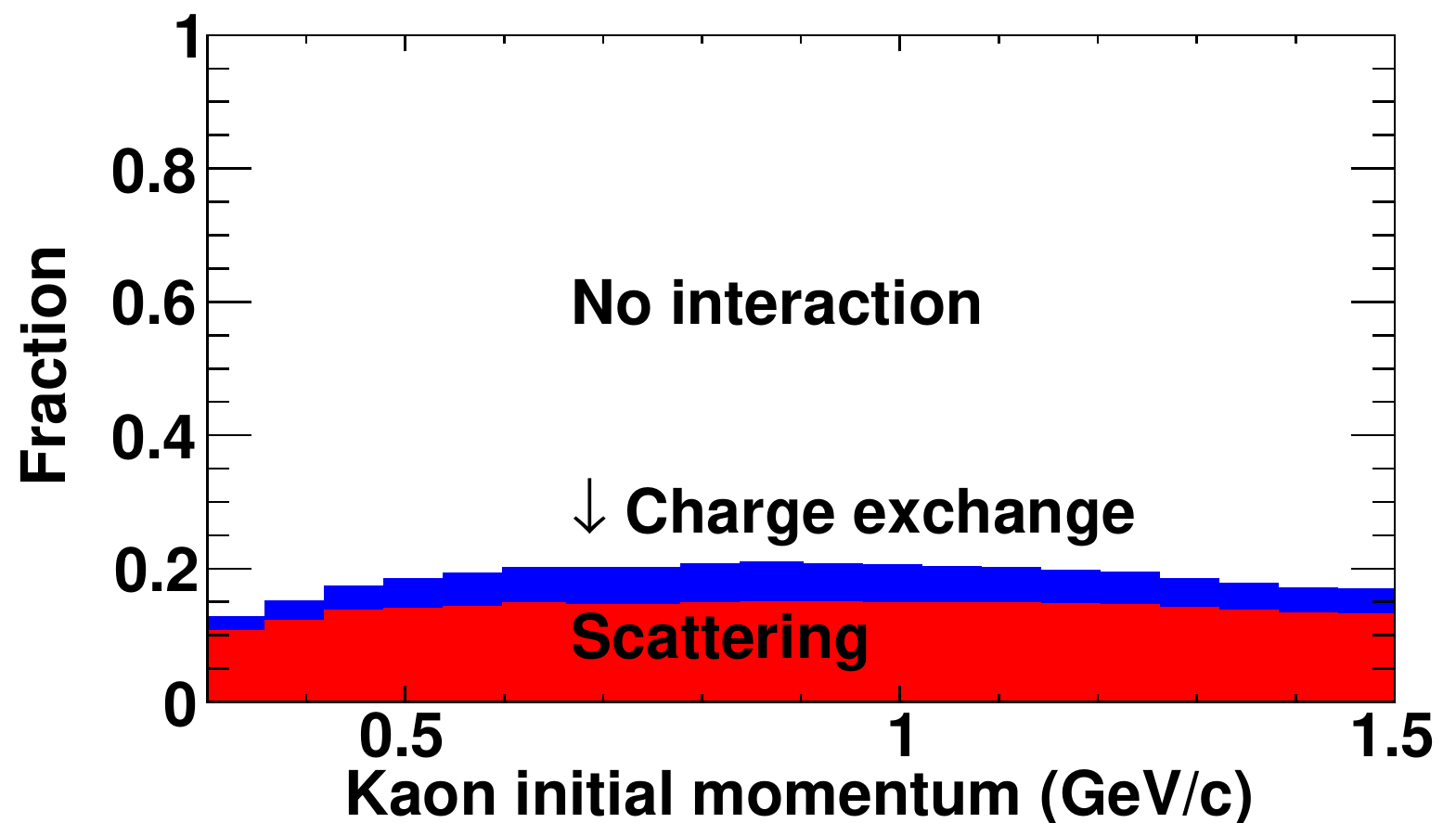}
        \end{minipage}
    \end{tabular}
    \caption{Fraction of $K^0$ interactions in the ${}^{16}\mathrm{O}$ nucleus as a function of $K^0$ momentum. Elastic scattering (red) and charge exchange (blue) are simulated in the nucleus.}
    \label{fig:K_scat_fraction}
\end{figure}

\begin{figure}[htbp]
    \centering
    \begin{tabular}{c}
        \begin{minipage}{1.0\hsize}
            \centering
            \includegraphics[keepaspectratio, width=1.0\textwidth]{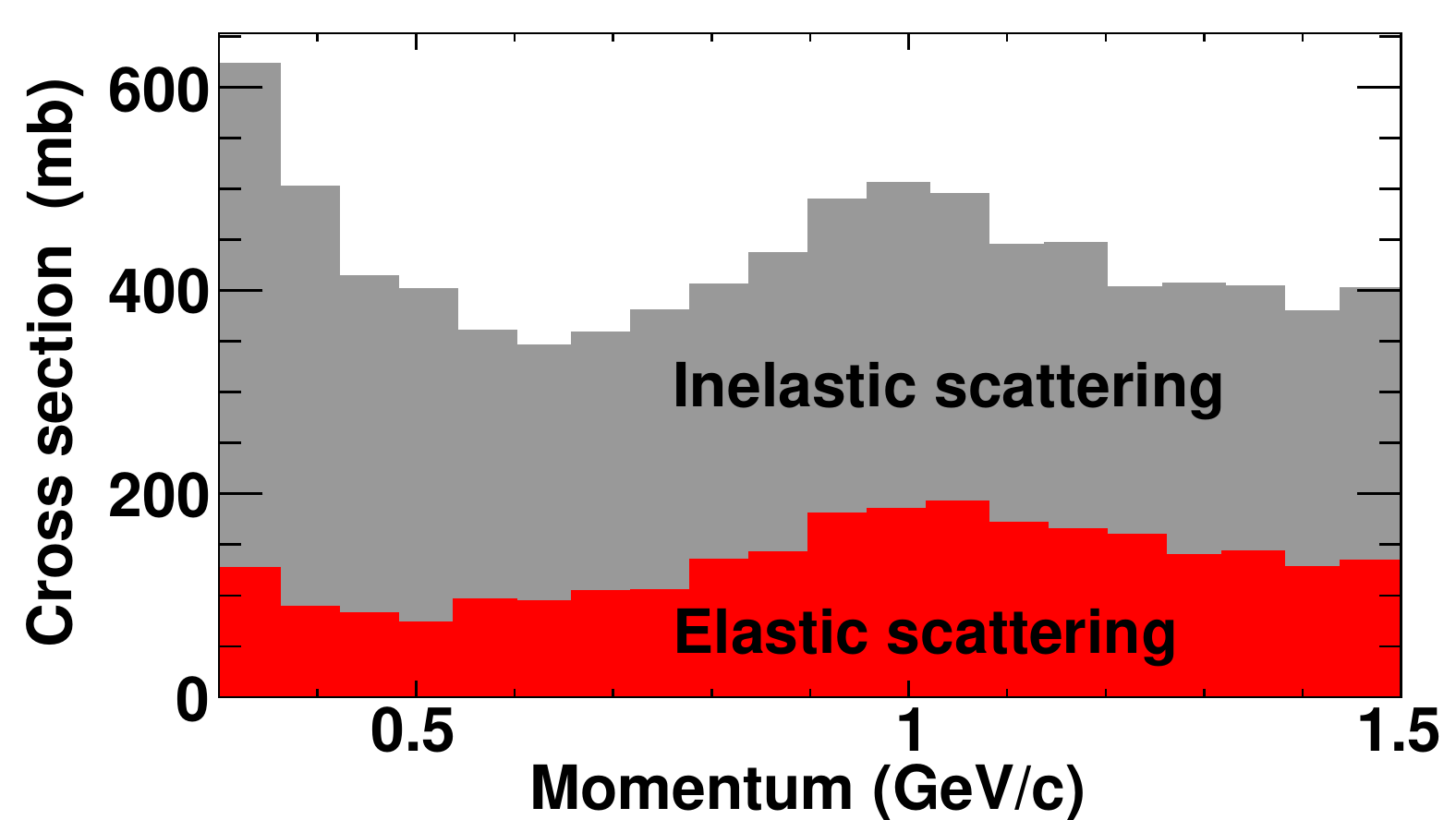}
        \end{minipage}
    \end{tabular}
    \caption{Cross section of $K^0_L$ in water. In the simulation of $K^0_L$ interactions, inelastic scattering (gray) and elastic scattering (red) are implemented.
    }
    \label{fig:Kaon_CS}
\end{figure}

\begin{figure}[htbp]
    \centering
    \begin{tabular}{c}
        \begin{minipage}{1.0\hsize}
            \centering
            \includegraphics[keepaspectratio, width=1.0\textwidth]{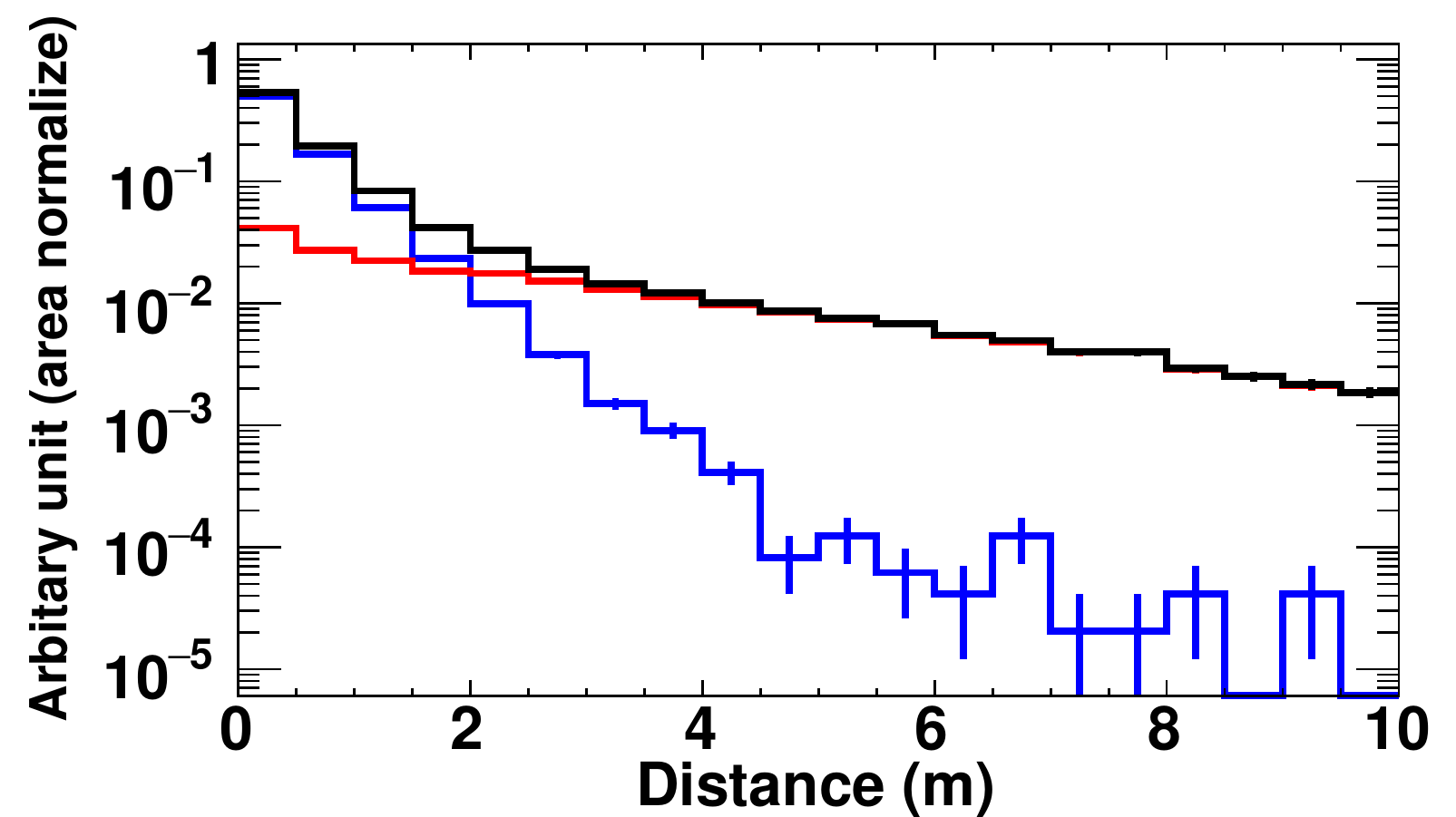}
        \end{minipage}
    \end{tabular}
    \caption{Distance between vertices of primary proton decay and $K^0_L$ decay.
    Total $p\to\mu^+K^0_L$ events (black) and the breakdown of $K^0_L$ decay events (red) and hadronic inelastic scattering events (blue) are shown.
    }
    \label{fig:Kdistance_true_log}
\end{figure}

Figure \ref{fig:K_scat_fraction} shows the fraction of $K^0$ interactions in the nucleus as a function of $K^0$ momentum.
After the kaon leaves the nucleus, interactions in water are simulated assuming eigenstates of $K^0_S$ and $K^0_L$.
Figure \ref{fig:Kaon_CS} shows the cross section of $K^0_L$ in the simulation.
$K^0_S$ and $K^0_L$ cross sections are calculated from the $K^0$ and $\bar{K}^0$ scattering amplitudes.
The $K^0N$ scattering amplitude is derived from the $K^+N$ cross section \cite{Hyslop_Kaon_CS} as written above while the $\bar{K}^0N$ scattering amplitude is calculated from the $K^-N$ cross section \cite{Martin_Kaon_CS}.
Figure \ref{fig:Kdistance_true_log} shows the distance between the positions of proton decay and $K^0_L$ decay.
The mean free path calculated from the cross section is about 1\,m in water while the average travel length of $K^0_L$ with 326.5 MeV/c momentum is 13\,m if it is in vacuum.
Therefore, $K^0_L$ from proton decay scatters in water before it decays in most cases.
The flight length becomes shorter if hadronic inelastic scattering occurs.

In contrast, $K^0_S$ immediately decays in water.
Coherent regeneration from $K^0_L$ to $K^0_S$ is also implemented both in the oxygen nucleus and in water.
The regeneration probability in water is based on the results of a kaon scattering experiment using a carbon target \cite{Eberhard_regen}.
In the oxygen nucleus, the regeneration probability is assumed to be proportional to its density relative to water.
The fraction with regeneration is about 0.1\% of the total $p\to\mu^+K^0_L$ MC events.

\subsection{Atmospheric neutrinos}
Atmospheric neutrinos are the dominant source of background in the proton decay search.
Among the atmospheric neutrino interactions, major components of the background events consist of pion production like single pion production or deep inelastic scattering since the signal events have pions from kaon decay.
The contribution of kaon production is relatively small due to the small cross section.
The atmospheric neutrino flux is calculated by the model of M. Honda {\it et al.} \cite{Honda1}\cite{Honda2}.
Neutrino interactions in the water are simulated by the NEUT program \cite{NEUT}.
The effect of two-flavor neutrino oscillation is accounted in the background estimation assuming $\sin^22\theta=1.0$ and $\Delta m^2= 2.5\times10^{-3}\,\mathrm{eV^2}$ which are the same parameters as \cite{Regis}, while the variation of the background rates is negligible with three-flavor oscillation \cite{PDG}.

\section{Reconstruction}
\label{sec:reconstruction}
Using event reconstruction we estimate basic event properties such as vertex, particle type and momentum based on PMT hit information.
Cherenkov rings are classified as showering (electron-like) or non-showering (muon-like)
Electrons and gammas are identified as the showering type because they produce electromagnetic showers and generate diffuse Cherenkov rings, while muons and charged pions do not produce showers and generate Cherenkov rings with clear edges.
In the previously published SK-I to SK-III search, the vertex position is reconstructed first by the timing information.
Then, the number of rings is determined and the particle type and momentum are reconstructed from the observed photoelectrons and pattern of PMT hits in separate steps.
In this study, the improved event reconstruction algorithm fiTQun \cite{fiTQun_MiniBooNE} is used.
It is based on the maximum likelihood method for both hit and unhit PMTs.
The vertex, momentum and particle type of the ring are simultaneously determined from the time and hit pattern information in fiTQun.
The performance of reconstruction is significantly improved with fiTQun \cite{Miao_atm}.

$K^0_L$'s produced from proton decay undergo hadronic interactions while propagating in the water, lose kinetic energy and decay a few meters away with a lifetime of 51\,ns.
In fiTQun, multiple rings are searched for assuming a single vertex in the initial step of the reconstruction.
This leads to the degradation of the reconstruction performance for charged particles generated at displaced vertices.
For this reason, fiTQun was modified for $p\to\mu^+K^0_L$ event selection to reconstruct a muon from proton decay and particles from kaon decay.
After the first ring is reconstructed, the hit PMTs belonging to that ring are masked and the second ring is searched for different vertex using the remaining hits.
Masked PMTs are not used for the calculation of the likelihood of the second ring.

As a further update from the previous study, a neutron tagging algorithm \cite{Irvine_ntag} is used for the $p\to\mu^+K^0$ search. 
Neutrons generated in water are captured by the surrounding hydrogen ($n+p\to d+\gamma$) after about 200\,$\mu$s in average. 
Neutrons can be identified by detecting the 2.2\,MeV gamma rays emitted by this capture.
Until SK-III, only the hits within 1.3\,$\mu$s around the trigger were recorded, so the neutrons could not be identified. 
Since SK-IV, the electronics modules were updated to record all PMT hits from 35\,$\mu$s to 535\,$\mu$s after the primary trigger.
This enables detection of neutron capture on hydrogen in that time window.
The detection efficiency of neutrons is about 20\% in the SK-IV period because the 2.2\,MeV gamma rays are easily buried by the dark hits of PMT and environmental radioactivity which mainly produces events below 4\,MeV \cite{Nakano_radon}.
Neutrons are abundantly generated by atmospheric neutrino interactions and by the secondary interactions of the hadrons in water.
On the other hand, neutrons are generated with about 10\% probability after proton decay in oxygen due to de-excitation of ${}^{15}\mathrm{N}$ \cite{Ejiri}.
As a result, neutrino interactions have multiple neutrons but proton decay has at most one neutron and usually zero.

\section{Event selection}
\label{sec:selection}
There are five selections to extract each $K^0$ decay mode, two for $K^0_S$ decay ($K^0_S\to2\pi^0$, $K^0_S\to\pi^+\pi^-$) and three for $K^0_L$ decay ($K^0_L\to\pi^\pm l^\mp\nu_l$ where $l$ is an electron or muon, $K^0_L\to3\pi^0$, $K^0_L\to\pi^+\pi^-\pi^0$).
There are two updates to selection criteria from the previous analysis \cite{Regis}.
In the previous study, events with two and three Cherenkov rings were selected as candidates for the $K^0_S\to\pi^+\pi^-$ decay mode, while only three-ring events are selected in this study, as the background rate for two-ring events becomes higher than for the other selections.
Three selection criteria are applied for each $K^0_L$ decay mode instead of the single selection criterion which was used in the previous paper.
In all selections, we require that the events should be fully contained with the vertex in the fiducial volume defined as a region more than 2\,m away from the ID wall (FCFV selection).
The fully contained events are selected by the condition that the number of hit PMTs in the largest OD hit cluster should be less than 16 hits.
In addition, we require that the total visible energy (electron equivalent total energy deposit in the detector) should be greater than 30\,MeV ($E_\mathrm{vis}$ selection).
In the recent $p\to e^+\pi^0$ and $p\to\mu^+\pi^0$ searches \cite{SK_epi0_3}, an enlarged fiducial volume cut (vertex is more than 1\,m away from the ID wall) was used for the proton decay search.
It was not employed in this analysis because the $K^0_L$ decays a few meters away from the point of the proton decay as shown in Figure \ref{fig:Kdistance_true_log}, and the $K^0_L$ decay point could be close to the ID wall or outside the ID if the proton decays within 2\,m of the ID wall.
The final samples of the selection A (A and B) are excluded in the selection B (C).

\subsection{Selection for $p \to \mu^+K^0_S,\, K^0_S\to2\pi^0$}
\begin{figure*}[htbp]
    \centering
    \begin{tabular}{ccc}
        \begin{minipage}{0.33\hsize}
            \centering
            \includegraphics[keepaspectratio, width=1.0\textwidth]{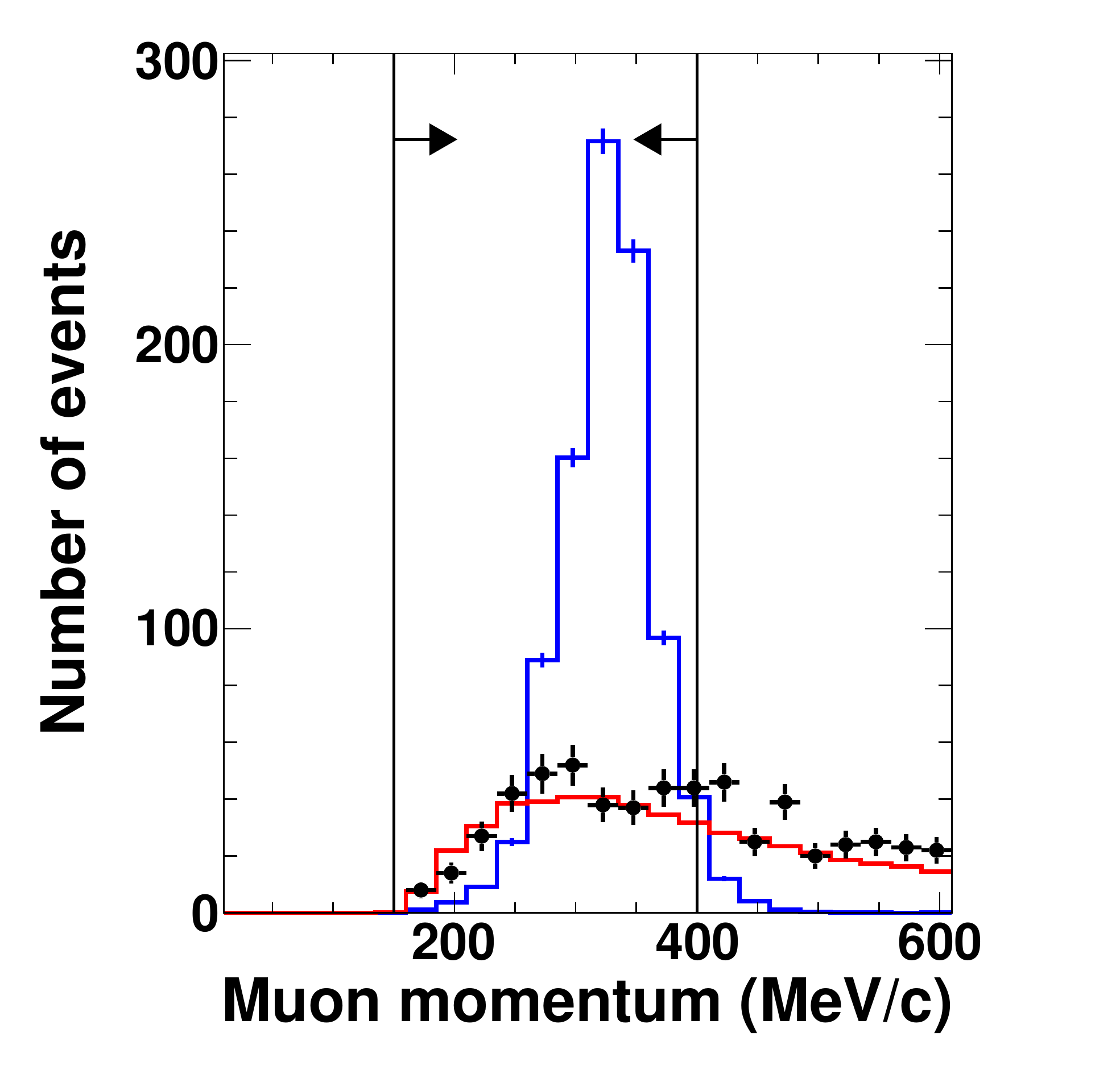}
        \end{minipage}
        &
        \begin{minipage}{0.33\hsize}
            \centering
            \includegraphics[keepaspectratio, width=1.0\textwidth]{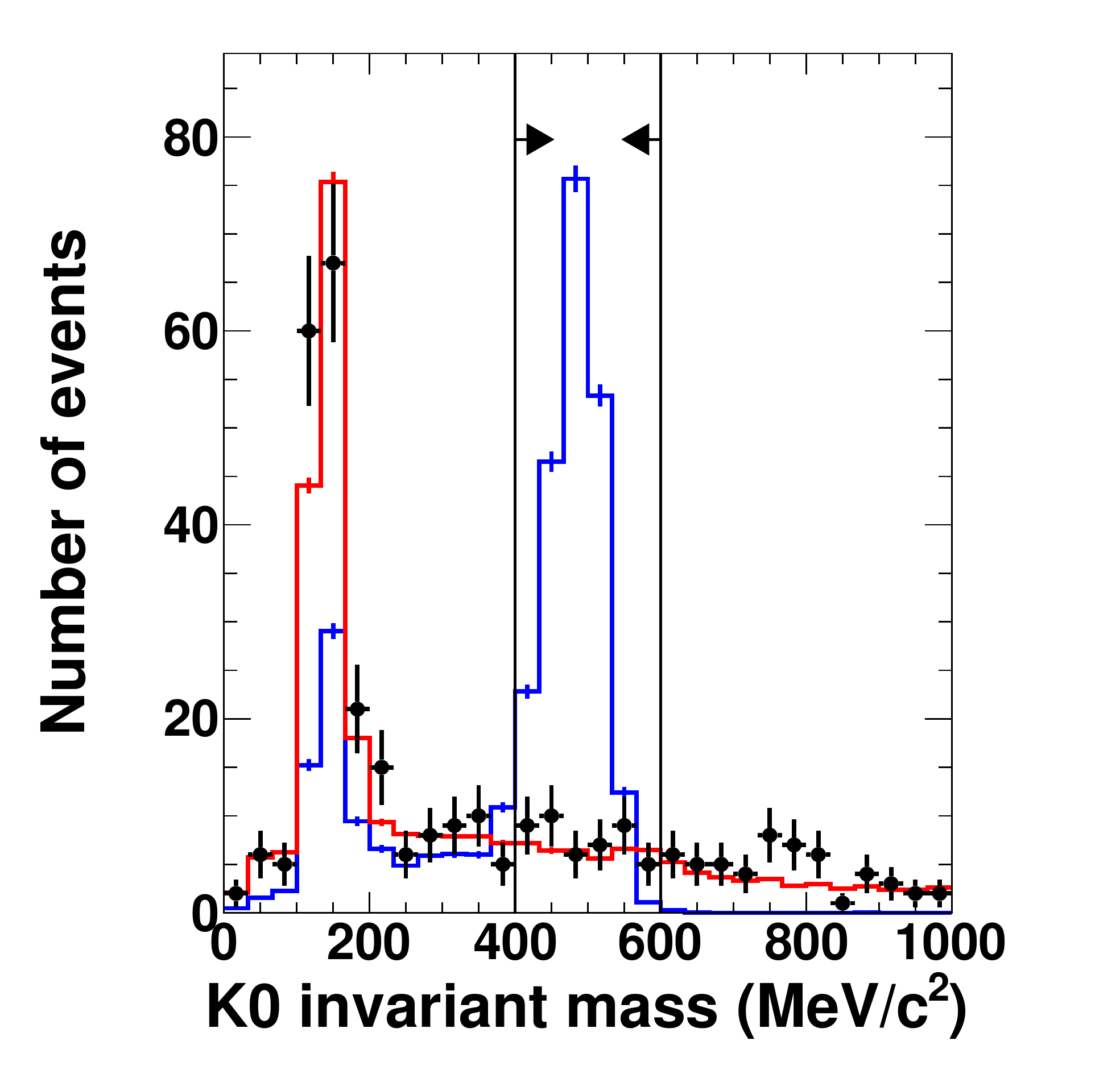}
        \end{minipage}
        &
        \begin{minipage}{0.33\hsize}
            \centering
            \includegraphics[keepaspectratio, width=1.0\textwidth]{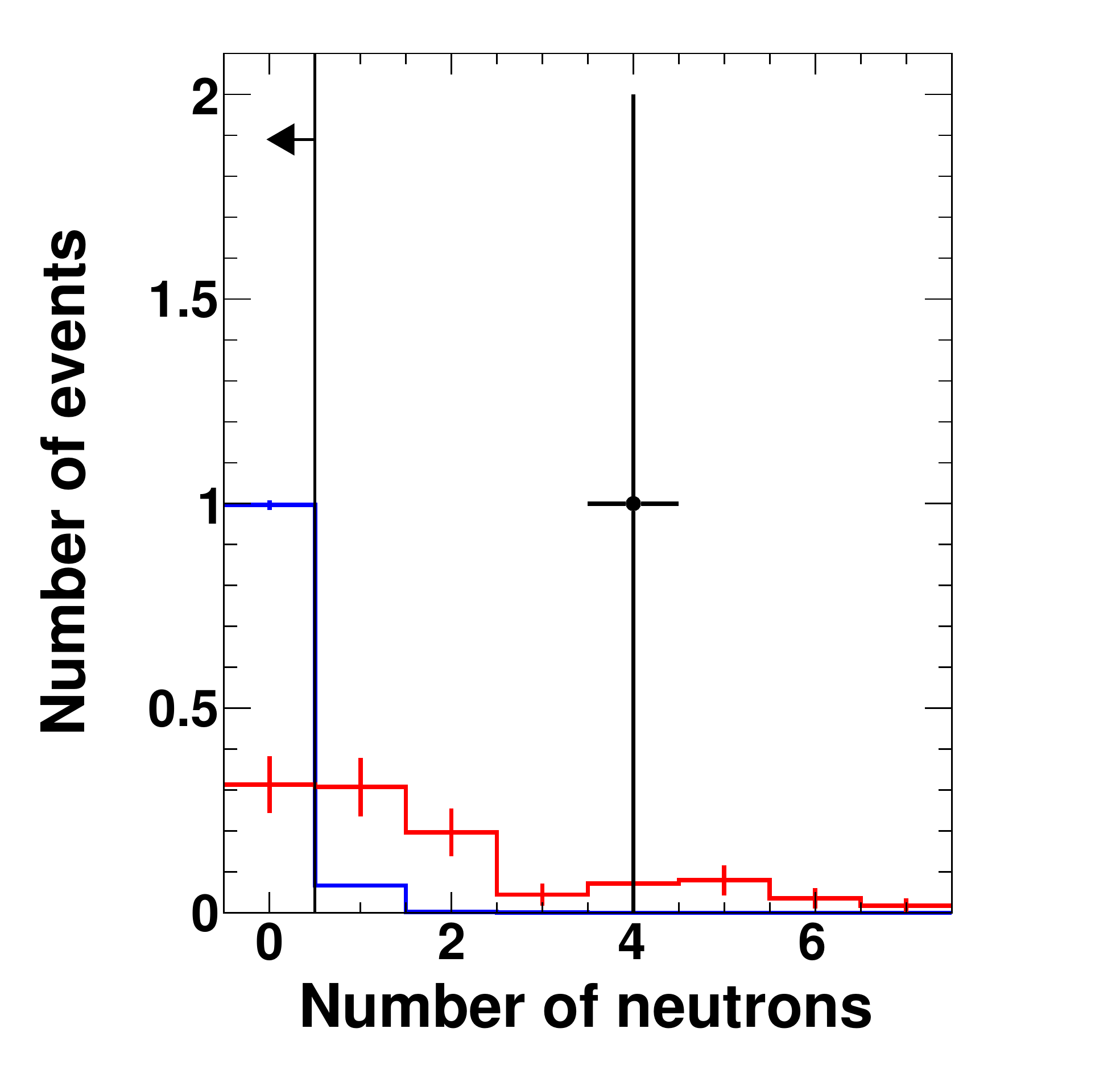}
        \end{minipage}
    \end{tabular}
    \caption{Reconstructed muon momentum after selection A-4 (left), Kaon invariant mass after selection A-5 (center) and number of tagged neutrons after selection A-8 (right).
    The data (black dots) are compared with the atmospheric neutrino MC (red) normalized by the livetime, and the signal MC (blue) normalized to the atmospheric neutrino MC.
    }
    \label{fig:mom_mu_mass_K_ntag_nn_sel0}
\end{figure*}

\begin{figure*}[htbp]
    \centering
    \includegraphics[keepaspectratio, width=0.8\hsize]{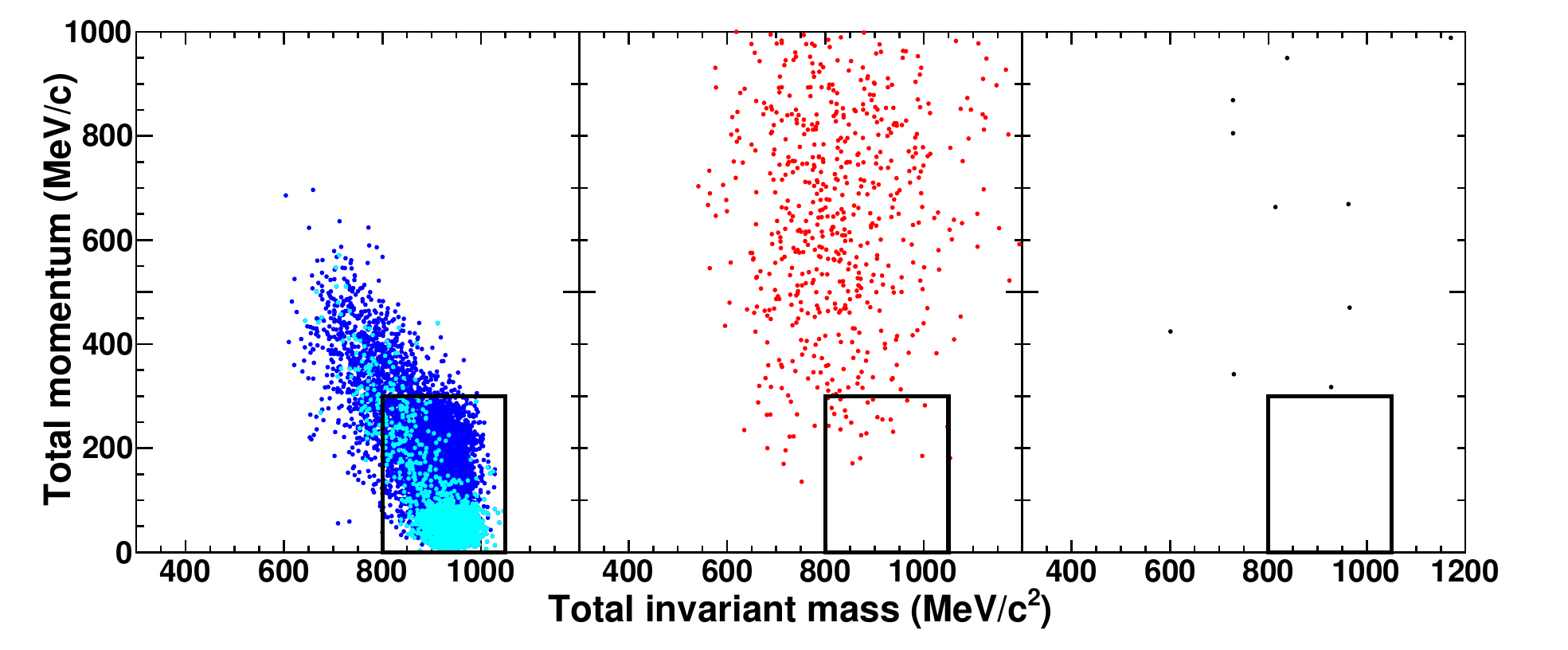}
    \caption{
        Scatter plot of the reconstructed total invariant mass and total momentum after applying all cuts in $K^0_S\to2\pi^0$ selection except those on the plotted variables. 
        From left to right, signal MC, atmospheric neutrino MC (500\,years-equivalent) and data (3244.39 live-days) are shown.
        In signal MC, cyan shows free protons and blue shows bound protons.
        }
    \label{fig:2d_sel0}
\end{figure*}

The events which satisfy the following criteria are selected as candidates for $K^0_S\to2\pi^0$:
\begin{itemize}
    \item[A-1:] Events should pass FCFV and $E_\mathrm{vis}$ selection.
    \item[A-2:] The number of rings should be three, four or five.
    \item[A-3:] There must be one non-showering ring. It is assumed as the primary muon.
    \item[A-4:] There must be one Michel electron.
    \item[A-5:] The reconstructed momentum of the non-showering ring should be $150 < P_\mu < 400\,\mathrm{MeV}/c$. 
    \item[A-6:] The reconstructed invariant mass of the showering rings should be $400 < M_K <600\,\mathrm{MeV}/c^2$.
    \item[A-7:] The reconstructed total momentum should be $P_\mathrm{tot} < 300\,\mathrm{MeV}/c$.
    \item[A-8:] The reconstructed total invariant mass should be $ 800 < M_\mathrm{tot} < 1050\,\mathrm{MeV}/c^2$.
    \item[A-9:] There should be no tagged neutrons.
\end{itemize}
Figure \ref{fig:mom_mu_mass_K_ntag_nn_sel0} shows muon momentum after selection A-4, kaon invariant mass after selection A-5 and the number of neutrons after selection A-8.
A peak around 150\,MeV$/c^2$ in the kaon invariant mass distribution is due to neutral pions either from atmospheric neutrino interactions or from decay of kaons from proton decay.
For this decay mode, 93\% of the signal events have no neutrons, while background events tend to have tagged neutrons.
A candidate event remaining after selection A-8 was rejected by requiring no tagged neutrons.
Figure \ref{fig:2d_sel0} shows scatter plots of the reconstructed total invariant mass and total momentum after applying all the cuts except those on the plotted variables.
Free proton events in the signal MC which appear around the top left corner of the signal box are due to secondary $K^0_S$ decay generated by $K^0_L$ scattering.

\subsection{Selection for $p \to \mu^+K^0_S,\, K^0_S\to\pi^+\pi^-$}
\begin{figure}[htbp]
    \centering
    \begin{tabular}{c}
        \begin{minipage}{1.0\hsize}
            \centering
            \includegraphics[keepaspectratio, width=1.0\hsize]{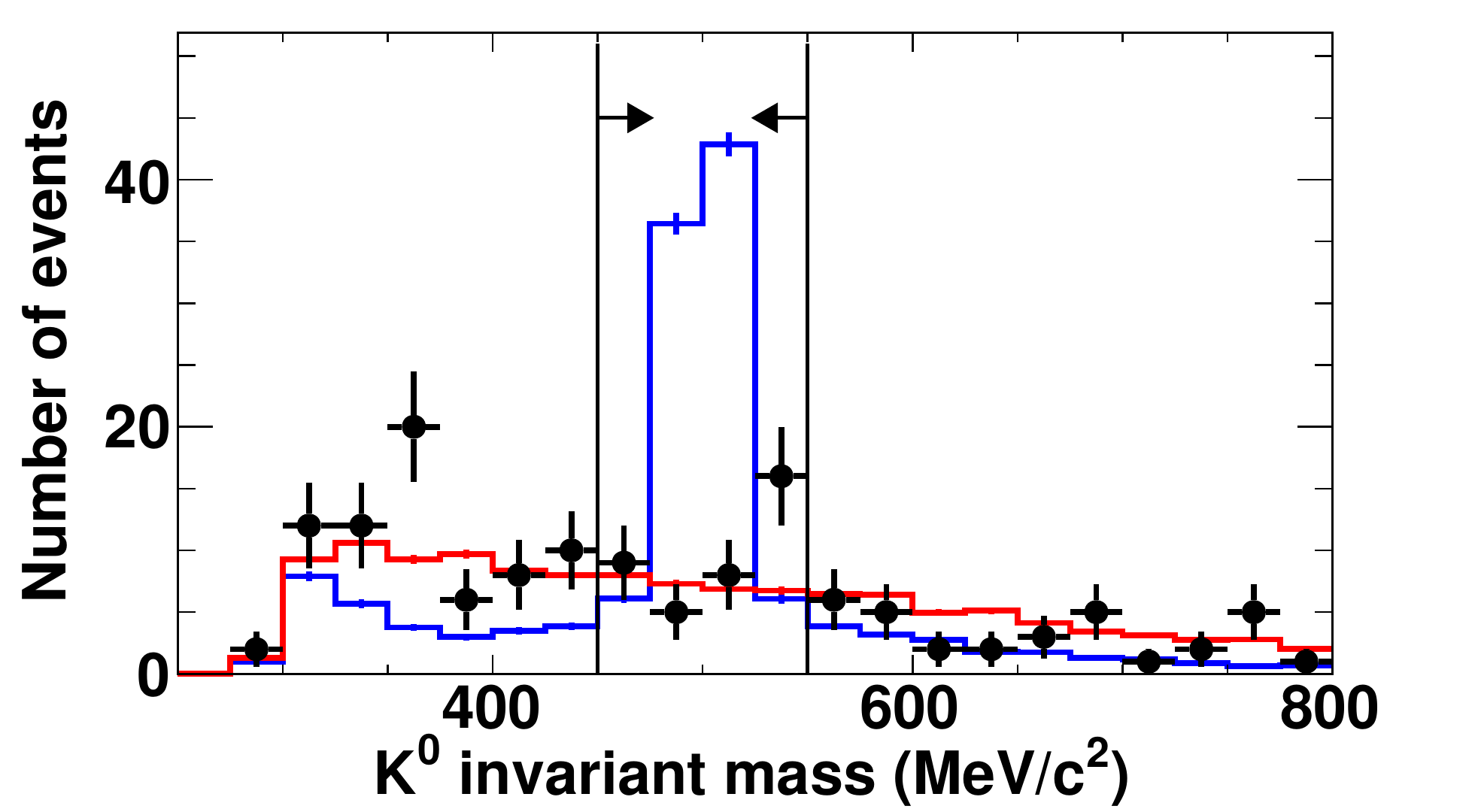}
        \end{minipage}
    \end{tabular}
    \caption{Reconstructed invariant mass distribution for kaons after applying $K^0_S\to\pi^+\pi^-$ selection B-4.
    The data (black dots) are compared with the atmospheric neutrino MC (red) normalized by the livetime, and the signal MC (blue) normalized to the atmospheric neutrino MC.
    }
    \label{fig:mass_K_sel2}
\end{figure}

\begin{figure*}[htbp]
    \centering
    \includegraphics[keepaspectratio, width=0.8\hsize]{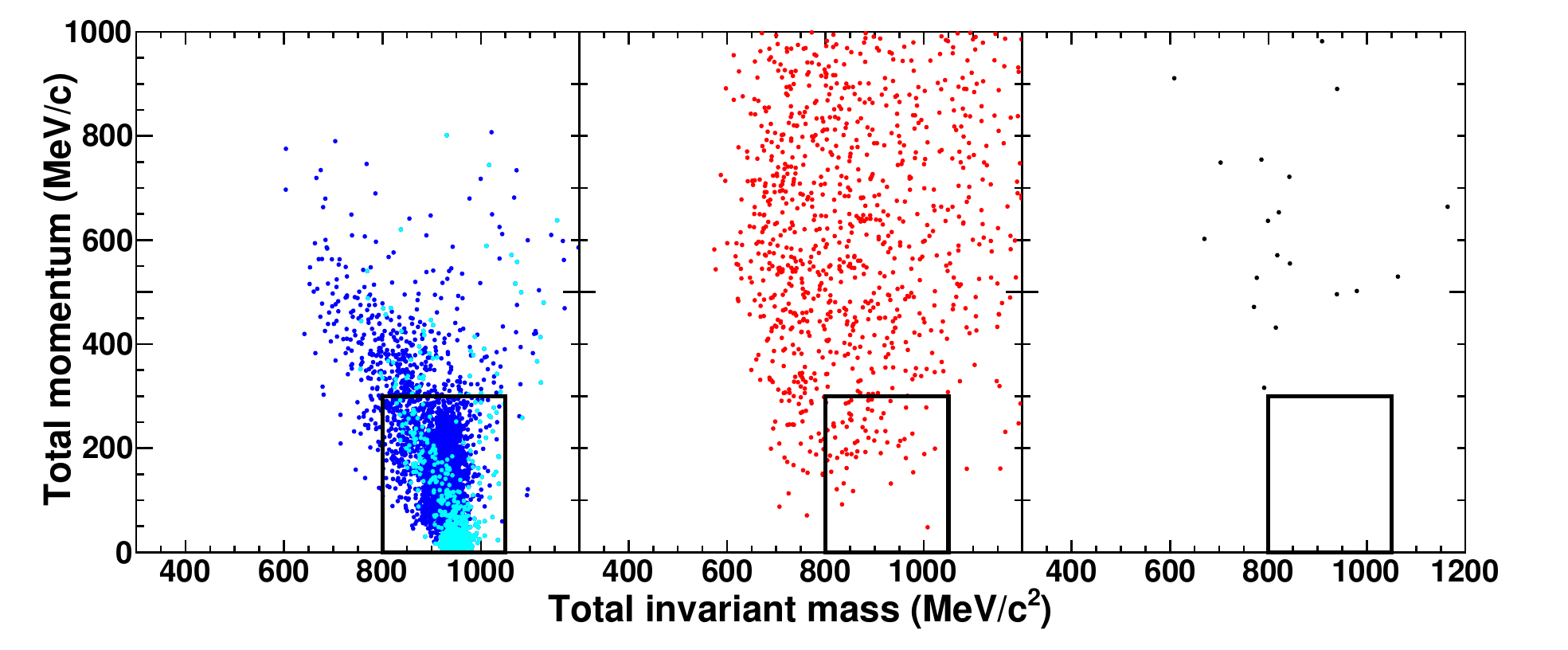}
    \caption{
        Scatter plot of the reconstructed total invariant mass and total momentum after applying all cuts in $K^0_S\to\pi^+\pi^-$ selection except those on the plotted variables. 
        From left to right, signal MC, atmospheric neutrino MC (500\,years-equivalent) and data (3244.39 live-days) are shown.
        In signal MC, cyan shows free protons and blue shows bound protons.
        }
    \label{fig:2d_sel2}
\end{figure*}

The events which satisfy the following criteria are selected as candidates for $K^0_S\to\pi^+\pi^-$:
\begin{itemize}
    \item[B-1:] Events should pass FCFV and $E_\mathrm{vis}$ selection.
    \item[B-2:] The number of rings should be three.
    \item[B-3:] All rings should be non-showering.
    \item[B-4:] The number of Michel electrons should be one or two.
    \item[B-5:] The reconstructed invariant mass of the second and third energetic non-showering rings should be $450 < M_K <550\,\mathrm{MeV}/c^2$.
    \item[B-6:] The reconstructed total momentum should be $P_\mathrm{tot} < 300\,\mathrm{MeV}/c$.
    \item[B-7:] The reconstructed total invariant mass should be $800 < M_\mathrm{tot} < 1050\,\mathrm{MeV/c^2}$.
\end{itemize}
In this selection, the most energetic ring is recognized as the muon from primary proton decay. 
The other rings are taken to be charged pions from $K^0_S$ decay.
$\pi^-$ from $K_S^0$ decay are often captured on a ${}^{16}\mathrm{O}$ nucleus and the number of tagged decay electrons in the signal event is expected to be one or two, one from the primary muon and the other from $\pi^+$.
No neutron tagging criterion is applied because about 45\% of signal events have tagged neutrons from pion capture.
Figure \ref{fig:mass_K_sel2} shows kaon invariant mass after the selection B-4.
The lower edge of the distribution is due to the Cherenkov threshold of charged pions.
Figure \ref{fig:2d_sel2} shows scatter plots of the reconstructed total invariant mass and total momentum after applying all cuts except those on the plotted variables.

\subsection{Selections for $p \to \mu^+K^0_L$}
\begin{figure}[htbp]
    \centering
    \begin{tabular}{c}
        \begin{minipage}{1.0\hsize}
            \centering
            \includegraphics[keepaspectratio, width=1.0\hsize]{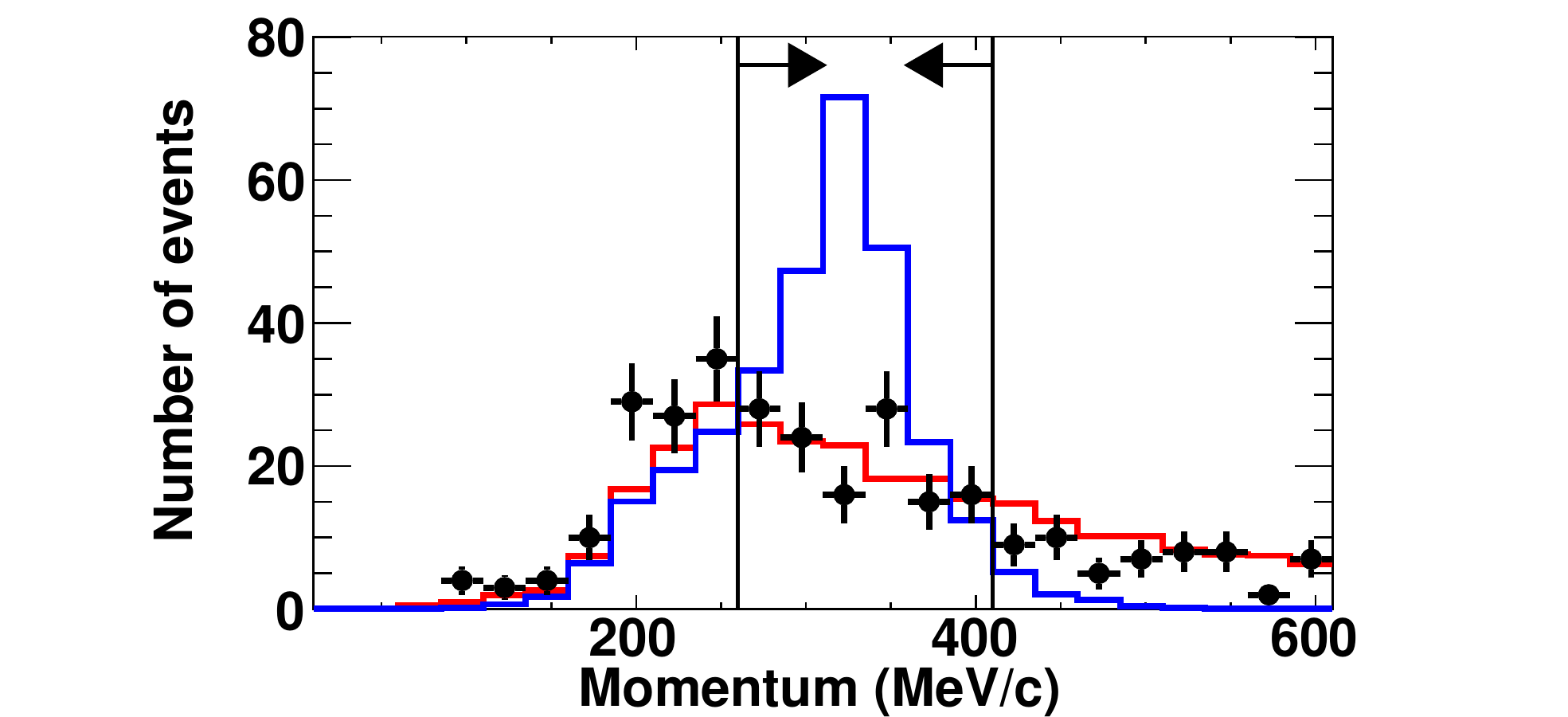}
        \end{minipage}
    \end{tabular}
    \caption{Reconstructed muon momentum of all events after C-5-1, C-5-2 and C-5-3 selections. The data (black dots) are compared with the atmospheric neutrino MC (red) normalized by the livetime, and the signal MC (blue) normalized to the atmospheric neutrino MC.}
    \label{fig:mom_mu_marged}
\end{figure}

\begin{figure}[htbp]
    \centering
    \begin{tabular}{c}
        \begin{minipage}{1.0\hsize}
            \centering
            \includegraphics[keepaspectratio, width=1.0\hsize]{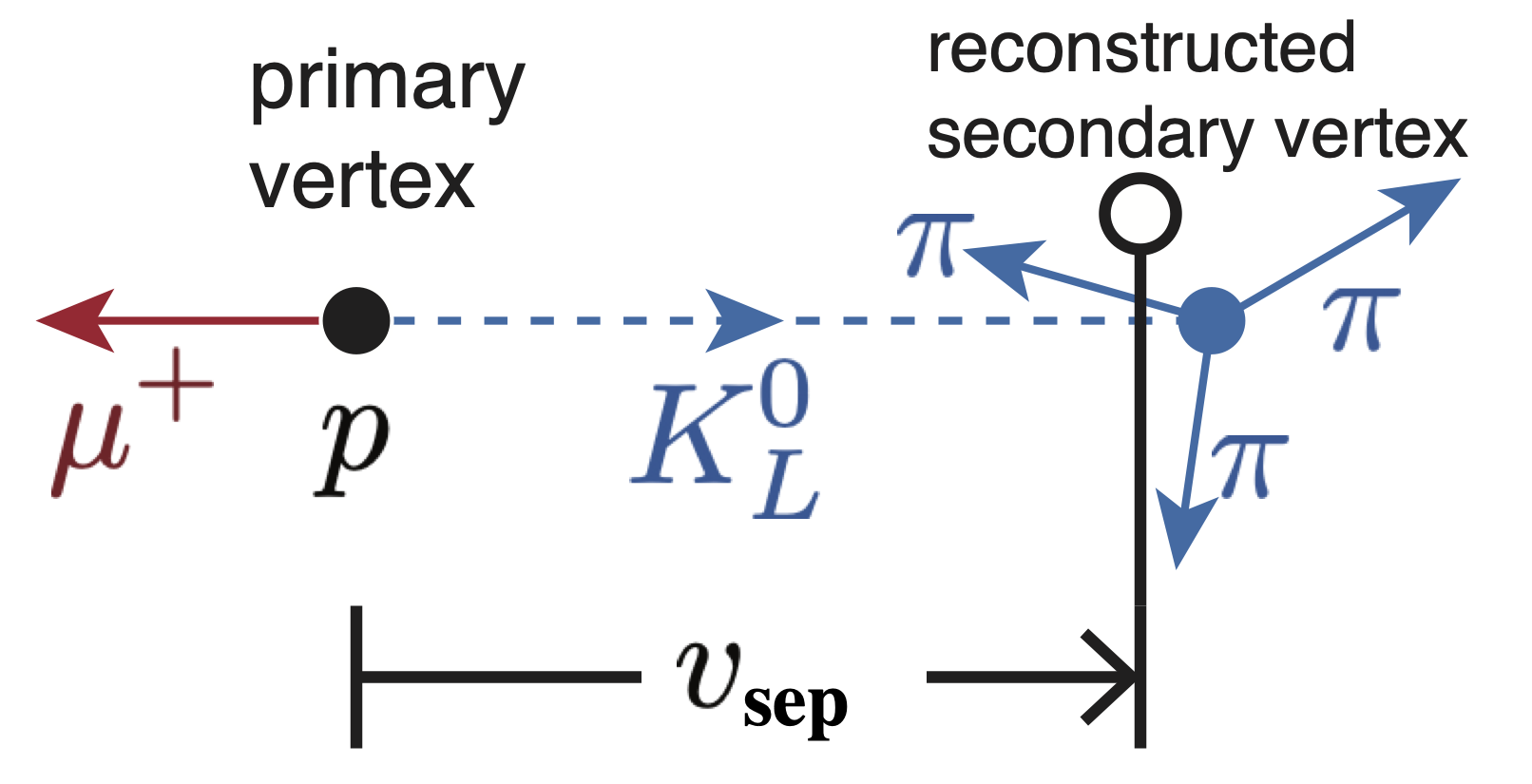}
        \end{minipage}
    \end{tabular}
    \caption{A schematic view of vertex separation.}
    \label{fig:def_of_vsep}
\end{figure}

\begin{figure}[htbp]
    \centering
    \begin{tabular}{c}
        \begin{minipage}{1.0\hsize}
            \centering
            \includegraphics[keepaspectratio, width=1.0\hsize]{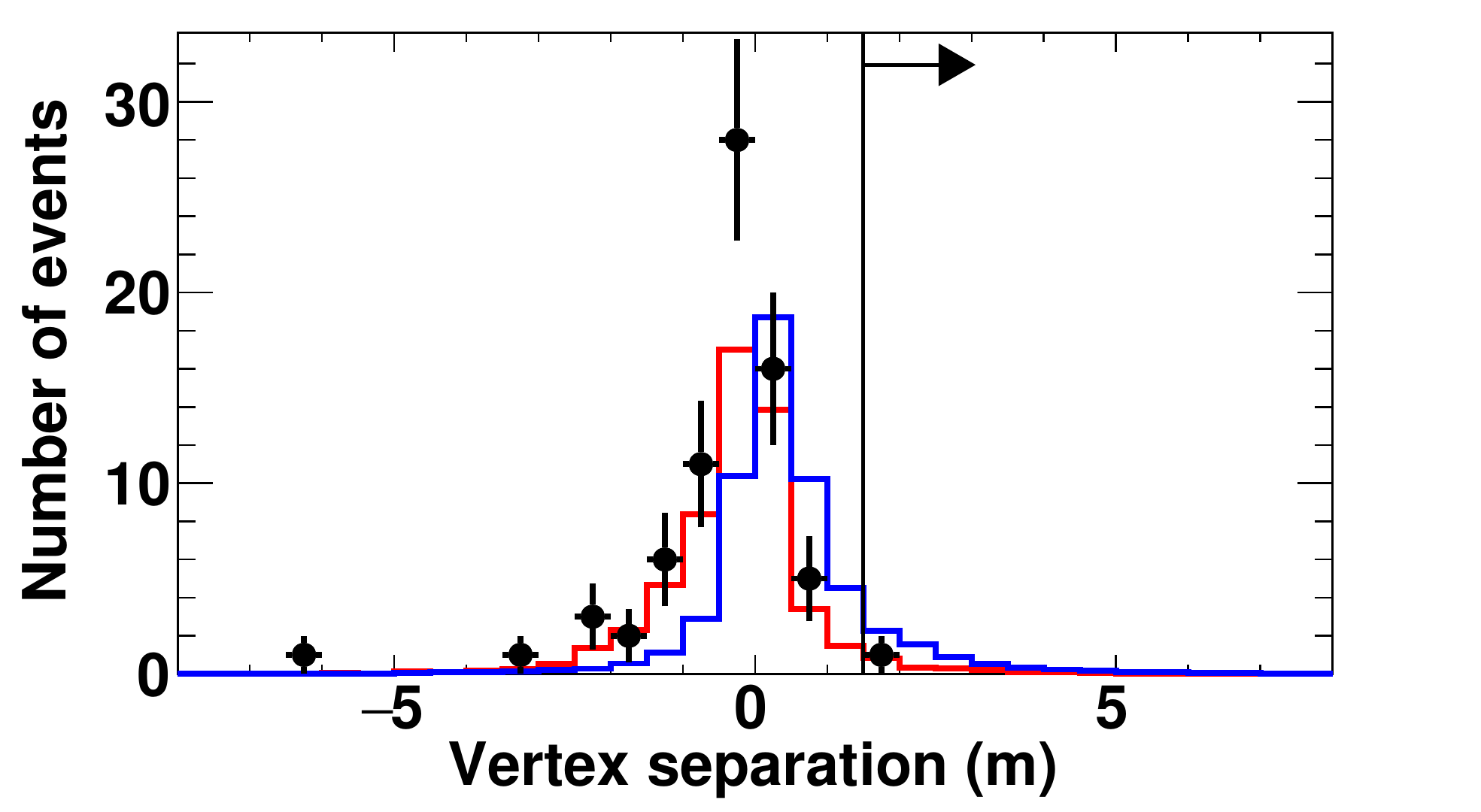}
        \end{minipage}
    \end{tabular}
    \caption{Reconstructed vertex separation distribution of all events in $K^0_L\to\pi^\pm l^\mp \nu$, $K^0_L\to3\pi^0$ and $K^0_L\to\pi^+\pi^-\pi^0$ selections. All cuts except vertex separation are applied. 
    The data (black dots) are compared with the atmospheric neutrino MC (red) normalized by the livetime, and the signal MC (blue) normalized to the atmospheric neutrino MC.
    }
    \label{fig:vsep}
\end{figure}

There are three selection criteria for $K^0_L$ decay, that is $K^0_L\to\pi^\pm l^\mp \nu$ ($l$ is electron or muon), $K^0_L \to 3\pi^0$ and $K^0_L \to \pi^+ \pi^- \pi^0$.
\begin{itemize}
    \item[C-1:] Events should pass FCFV and $E_\mathrm{vis}$ selection.
    \item[C-2:] Total observed photoelectrons (p.e.) should be $500 < Q_\mathrm{tot} < 8000\,\mathrm{p.e.}$
    \item[C-3:] \,
    \begin{itemize}
        \item[C-3-1:]The number of rings should be two or three (for $K^0_L\to\pi^\pm l^\mp \nu$).
        \item[C-3-2:]The number of rings should be four, five or six (for $K^0_L \to 3\pi^0$).
        \item[C-3-3:]The number of rings should be three or four (for $K^0_L \to \pi^+ \pi^- \pi^0$).
    \end{itemize} 
    \item[C-4:] \,
    \begin{itemize}
        \item[C-4-1:]The number of showering rings should be zero or one (for $K^0_L\to\pi^\pm l^\mp \nu$).
        \item[C-4-2:]The number of non-showering rings should be one (for $K^0_L \to 3\pi^0$).
        \item[C-4-3:]The number of showering rings should be two (for $K^0_L \to \pi^+ \pi^- \pi^0$).
    \end{itemize} 
    \item[C-5:] \,
    \begin{itemize}
        \item[C-5-1:] The number of Michel electrons should be two or three (for $K^0_L\to\pi^\pm l^\mp \nu$).
        \item[C-5-2:]The number of Michel electrons should be one (for $K^0_L \to 3\pi^0$).
        \item[C-5-3:]The number of Michel electrons should be two or three (for $K^0_L \to \pi^+ \pi^- \pi^0$).
    \end{itemize} 
    \item[C-6:] The reconstructed muon momentum should be $260 < P_\mu < 410\,\mathrm{MeV}/c$ 
    \item[C-7:] The reconstructed vertex separation should be $1.5\,\mathrm{m}< v_\mathrm{sep}$ (defined in the text below).
    \item[C-8:] There should be no tagged neutrons.
\end{itemize}
In the $K^0_L$ decay selections, the primary non-showering ring is chosen as the primary muon from proton decay.
Figure \ref{fig:mom_mu_marged} shows the reconstructed muon momentum after C-5 selection.
Since these $K^0_L$ decays are three-body decays with relatively long lifetimes, it is difficult to reconstruct all secondaries.
Therefore strict cuts to the invariant mass are not applied in these selections. 
Instead, the vertex separation $v_\text{sep}$, defined as the distance between the primary (muon) and secondary (kaon) vertices along the opposite direction of the primary muon as illustrated in Figure \ref{fig:def_of_vsep}, is used to distinguish the signal events from the backgrounds.
Typical $p\to\mu^+K^0_L$ events have positive vertex separation due to opposite directions of muon and kaon from proton decay, while the vertex separation of typical atmospheric neutrino events is zero. 
Figure \ref{fig:vsep} shows vertex separation distributions after applying all cuts except for C-7.
The peak positions between the signal and atmospheric neutrino MC differ by about 0.5\,m.
The distribution of the atmospheric neutrino MC has a larger tail in the negative region.
This is mainly due to Michel electrons from muons produced by $\nu_\mu$ CC interaction and the scattering of charged pions in water which causes multiple Cherenkov rings.
The value of the vertex separation cut was chosen to optimize for the sensitivity of proton decay search.

\section{Results}
\label{sec:result}
\begin{figure*}[htbp]
    \centering
    \begin{tabular}{c}
        \begin{minipage}{0.9\hsize}
            \centering
            \includegraphics[keepaspectratio, width=1.0\textwidth]{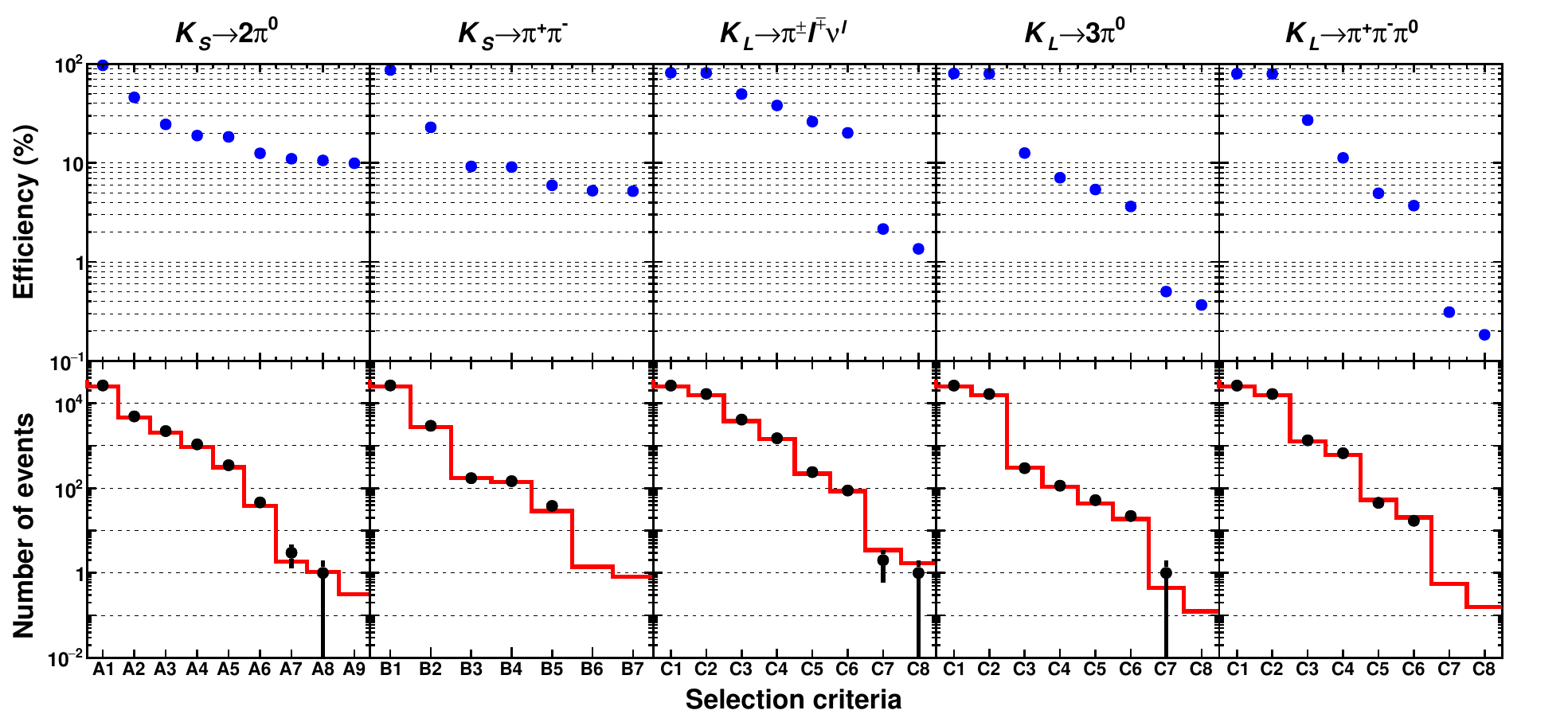}
        \end{minipage}
    \end{tabular}
    \caption{Signal efficiencies (blue), number of background events (red) and number of candidates (black) for each step of the selections for SK-IV. Error bars show statistical uncertainty. The number of atmospheric neutrino MC events is normalized by the livetime (3244.39\,live-days).}
    \label{fig:performance}
\end{figure*}

The results of these selections are summarized in Table \ref{tab:results}.
It gives the signal efficiency, the number of background events and the number of candidate events for each selection.
The efficiency of each selection is defined as a ratio of the number of signal events after selection to the number of generated events within 2 m from the ID wall.
The total efficiency is 17.0$\pm$1.2\% and the total number of expected background is $15.5\pm2.9$\,events/(Mton$\cdot$year). In the SK-IV period, 0.2\,Mton$\cdot$years were observed and $3.1\pm0.6$ background events are expected.
Systematic uncertainties are also shown for the signal efficiencies and the background rates.
The details of the major systematic uncertainties are explained in Section \ref{sec:syst}.
As a result of the selections, one candidate remains in the final samples for $K^0_L\to\pi^\pm l^\mp\nu$ decay modes. 
As there are no significant excesses beyond the expected backgrounds, limits of the proton lifetime are set in Section \ref{sec:lifetime}.
\begin{table*}[htbp]
    \centering
    \caption{Summary of the $p\to\mu^+K^0$ search. Uncertainties are quadratic sums of the MC statistical uncertainties and the systematic uncertainties. 
    The lower limit (SK-I$+$SK-II$+$SK-III$+$SK-IV  combined) is smaller than that from only SK-IV data due to two candidates of $K^0_S\to2\pi^0$ selection in SK-II data for the background expectation of 0.20 events \cite{Regis}.
    }
    \vspace{0.2cm}
    \begin{tabular}{ccccc} \hline\hline
      Search mode & Efficiency (\%) & Background (events) & Candidates (events) & Lower limit ($10^{33}$ years)\\ \hline
      $K^0_S\to2\pi^0$&$9.9\pm1.0$&$0.3\pm0.1$&0&2.8 \\
      $K^0_S\to\pi^+\pi^-$&$5.2\pm0.6$&$0.8\pm0.2$&0&1.5 \\
      $K^0_L\to\pi^\pm l^\mp\nu$&$1.4\pm0.3$&$1.7\pm0.5$&1&0.3 \\
      $K^0_L\to3\pi^0$&$0.37\pm0.05$&$0.12\pm0.06$&0&0.1 \\
      $K^0_L\to\pi^+\pi^-\pi^0$&$0.18\pm0.04$&$0.16\pm0.07$&0&0.05 \\
      \hline
      \multicolumn{4}{l}{SK-IV combined (199\,kton$\cdot$years)}&4.5\\
      \multicolumn{4}{l}{SK-I$+$SK-II$+$SK-III$+$SK-IV  combined (372\,kton$\cdot$years)} &3.6\\
      \hline \hline
    \end{tabular}
  \label{tab:results}
\end{table*}
Figure \ref{fig:performance} shows the signal efficiency, the number of background events and the number of candidates for each step of the selections.
For $K^0_S$ decay modes, about 95\% of background events are rejected by the total momentum cut while keeping about 90\% of the signal events.
For $K^0_L$ decay modes, more than 95\% of the remaining background events are rejected by the vertex separation cut whereas the signal efficiencies are about 10\%. 
Table \ref{tab:bg_breakdown} shows the breakdown of the remaining background events by the interaction mode.
\begin{table*}[htbp]
    \centering
    \caption{Breakdown of remaining atmospheric neutrino backgrounds by the interaction mode (\%). 
    CC, NC, QE and DIS stand for charged-current, neutral-current, quasi-elastic and deep inelastic scattering, respectively. Uncertainties are statistical.}
    \vspace{0.2cm}
    \begin{tabular}{cccccc} \hline\hline
        Modes & $K^0_S\to2\pi^0$ & $K^0_S\to\pi^+\pi^-$ & $K^0_L\to\pi^\pm l^\mp\nu$ & $K^0_L\to3\pi^0$ & $K^0_L\to\pi^+\pi^-\pi^0$ \\ \hline
        CCQE & $1.9\pm1.9$ & - & $9.7\pm2.8$ & $8.2\pm8.1$ & -\\
        CC1$\pi$ & $23.4\pm9.3$ & $70.2\pm6.3$ & $82.9\pm3.6$ & $28.9\pm15.4$ & $57.3\pm16.0$\\
        CC1$K$ & - & $3.4\pm2.5$ & - & - & -\\
        CC1$\eta$ & $37.3\pm10.6$ & - & - & - & -\\
        CCDIS & $31.6\pm11.5$ & $11.3\pm4.2$& $3.2\pm1.6$ & $48.6\pm17.7$ & $42.7\pm16.0$  \\
        NC & $5.8\pm5.6$ & $15.0\pm5.5$ & $4.2\pm2.0$ & $14.2\pm13.0$ & -\\
        \hline \hline
    \end{tabular}
  \label{tab:bg_breakdown}
\end{table*}
The major background source is not kaon production but single pion and eta production, and DIS, due to larger cross sections.

\section{Systematic uncertainty}
\label{sec:syst}

Tables \ref{tab:sys_sig} and \ref{tab:sys_bg} summarize the systematic uncertainties for the selection efficiencies and the expected background rates, respectively.
Uncertainties on the correlated decay probability \cite{Yamazaki_correlated_decay}, Fermi momentum models \cite{Nakamura_12Cscat}\cite{NEUT}, pion interaction \cite{Salcedo_pionFSI}\cite{Perio_pionFSI}\cite{Perio_PhD_pionFSI} and kaon interaction are considered in the signal efficiencies.
The kaon cross section in an oxygen nucleus was evaluated from the $K^0$-nucleon cross section accounting for the Fermi motion and Pauli blocking and the variation due to these effects is treated as the systematic uncertainty.
After leaving the nucleus, the kaon is simulated as eigenstates of $K^0_S$ and $K^0_L$.
The uncertainty of the $K^0_L$ cross section in water was evaluated from comparison of the simulation model and the independent measurement \cite{Cleland_K0scat}.
As the kaon interaction model is updated, systematic uncertainties for efficiencies of the SK-I to SK-III search were reevaluated to cover the difference between the old and new models.
They are summarized in Table \ref{tab:results_SK123} (Appendix).
Uncertainties in the neutrino flux, neutrino cross section and pion interaction were accounted in the estimation of backgrounds.
Among these, the neutrino cross section is the dominant source of systematic uncertainty.
As the systematic uncertainties associated with the detector performance and reconstruction, uncertainties in the number of events in the fiducial volume, detector non-uniformity, energy scale, ring counting, particle type identification, decay electron tagging and neutron tagging are considered in the signal efficiencies and background rates.
In addition, uncertainty in the vertex separation was evaluated from the deviation between the true and reconstructed vertex separations to be 11\% (12\%), 5\% (9\%) and 9\% (14\%) for the signal efficiencies (background rates) in $K^0_L\to\pi^\pm l^\mp\nu$, $K^0_L\to3\pi^0$ and $K^0_L\to\pi^+\pi^-\pi^0$ selections, respectively.
These are among the largest systematic uncertainties for the detector performance and reconstruction in $K^0_L$ selections.
\begin{table*}[htbp]
    \centering
    \caption{Systematic uncertainties on the signal efficiencies (\%)}
    \vspace{0.2cm}
    \begin{tabular}{cccccc} \hline\hline
        Sources & $K^0_S\to2\pi^0$ & $K^0_S\to\pi^+\pi^-$ & $K^0_L\to\pi^\pm l^\mp\nu$ & $K^0_L\to3\pi^0$ & $K^0_L\to\pi^+\pi^-\pi^0$ \\ \hline
        Correlated decay & 6.1 & 6.3 & 5.6 & 3.9 & 2.7 \\
        Fermi momentum & 1.0 & 1.8 & 1.3 & 0.1 & 1.0 \\
        Pion interaction & N/A & 4.5 & 2.7 & 1.8 & 6.3 \\
        Kaon interaction & 3.1 & 3.1 & 11.6 & 5.3 & 10.3 \\
        Reconstruction & 7.0 & 6.6 & 12.5 & 10.6 & 11.7 \\
        \hline
        Total & 9.9 & 10.8 & 18.2 & 12.6 & 17.0 \\
        \hline \hline
    \end{tabular}
  \label{tab:sys_sig}
\end{table*}
\begin{table*}[htbp]
    \centering
    \caption{Systematic uncertainties on the number of expected background events (\%)}
    \vspace{0.2cm}
    \begin{tabular}{cccccc} \hline\hline
        Sources & $K^0_S\to2\pi^0$ & $K^0_S\to\pi^+\pi^-$ & $K^0_L\to\pi^\pm l^\mp\nu$ & $K^0_L\to3\pi^0$ & $K^0_L\to\pi^+\pi^-\pi^0$ \\ \hline
        Neutrino flux & 8.7 & 6.8 & 6.6 & 8.0 & 7.9\\
        Neutrino interaction & 20.0 & 21.5 & 22.6 & 20.0 & 24.0 \\
        Pion interaction & 17.4 & 9.9 & 9.0 & 8.4 & 13.3 \\
        Reconstruction & 22.0 & 7.0 & 14.2 & 34.8 & 15.6 \\
        \hline
        Total & 35.5 & 25.6 & 28.9 & 41.4 & 32.5 \\
        \hline \hline
    \end{tabular}
  \label{tab:sys_bg}
\end{table*}

\section{Lifetime limit}
\label{sec:lifetime}
Since no statistically significant excesses were observed, a lower bound on the proton lifetime was calculated by using a Bayesian method \cite{Amsler_Bayes}\cite{Roe_Bayes}.
The probability distribution function of the decay width of proton decay is expressed by a Poisson distribution convolved with the systematic uncertainties as follows:
\begin{equation}
    \begin{split}
    P(\Gamma | n_i) = \iiint\frac{e^{-(\Gamma \lambda_i \epsilon_i+b_i)}(\Gamma\lambda_i \epsilon_i+b_i)^{n_i}}{n_i!} \\
    \times P(\Gamma)P(\lambda_i)P(\epsilon_i)P(b_i)d\epsilon_id\lambda_idb_i,
    \end{split}
\end{equation}
where $i$ is an index for each selection, $\Gamma$ is the decay rate, $n_i$ is the number of observed events, $\lambda_i$ is the exposure, $\epsilon_i$ is the signal efficiency and $b_i$ is the number of expected background events.
The probability distribution function of decay rate $P(\Gamma)$ is assumed to be uniform as there is no indication for its value.
$P(\lambda_i)$, $P(\epsilon_i)$ and $P(b_i)$ are the probability distribution functions for the exposure, efficiency and backgrounds, respectively:
\begin{equation}
    P(\lambda_i) \propto 
    \begin{cases}
    \exp{\left(\frac{-(\lambda_i-\mu_{\lambda_i})^2}{2\sigma^2_{\lambda_i}}\right)} & (\lambda_i > 0) \\
    0 & (\rm{otherwise})
    \end{cases}
\end{equation}
\begin{equation}
    P(\epsilon_i) \propto
    \begin{cases} 
    \exp{\left(\frac{-(\epsilon_i-\mu_{\epsilon_i})^2}{2\sigma^2_{\epsilon_i}}\right)} & (\epsilon_i > 0) \\
    0 & (\rm{otherwise})
    \end{cases}
\end{equation}
\begin{equation}
        P(b_i) \propto 
    \begin{cases}
        \int_{0}^{\infty}\frac{e^{-B}B^{n_{b_i}}}{n_{b_i}!}\exp{\left(\frac{-(C_{i}b_i-B)^2}{2\sigma^2_{b_i}}\right)} dB & (b_i > 0) \\
        0 & (\rm{otherwise})
    \end{cases}
\end{equation}
where $\sigma_{\lambda_i}$, $\sigma_{\epsilon_i}$ and $\sigma_{b_i}$ are the systematic uncertainties and $\mu_{\lambda_i}$ and $\mu_{\epsilon_i}$ are the expected exposure and efficiency, respectively.
$n_{b_i}$ is the number of expected backgrounds without livetime normalization and $C_i$ is a factor to normalize MC to the livetime of the data. 
The proton decay rate $\Gamma_{\rm limit}$ at a 90\% C.L. satisfies the following equation using this probability distribution function,
\begin{equation}
    0.9 = \int^{\Gamma_{\rm limit}}_0 d\Gamma \prod_{i}P(\Gamma | n_i).
\end{equation}
The lower limit of proton lifetime is expressed as the inverse of the decay rate:
\begin{equation}
    \tau_{\rm limit}/Br = 1/\Gamma_{\rm limit},
\end{equation}
where $Br$ is the branching ratio of the proton decay mode.
The lifetime limits by this method are also summarized in the Table \ref{tab:results}.
The limit from SK-IV data is $4.5\times10^{33}$\,years at 90\% C.L.

The limit from SK-I to SK-IV data was also evaluated.
For SK-I to SK-III, all channels in Table \ref{tab:results_SK123} were used.
The signal and background estimations for each period and decay mode were added as independent terms in Eq. (1) to (4) and combined in Eq. (5). 
Systematic uncertainties were assumed to be fully correlated for the entire period from SK-I to SK-IV assuming common sources.
As a result, the limit of $3.6\times10^{33}$\,years at 90\% C.L. was obtained from 0.37\,Mton$\cdot$years of data.
This limit is more than twice as long as the previous result, $1.6\times10^{33}$\,years, which uses data for SK-I, SK-II and SK-III. 
As explained in Section \ref{sec:syst}, the systematic uncertainties on the kaon scattering for SK-I, -II and III periods were reevaluated to cover the update of the kaon interaction model when the combined lifetime limit was calculated.
This affects mainly the signal efficiency of $K^0_L$ selection.
The lifetime limit for SK-I, -II and III becomes $1.2\times10^{33}$\,years with this update. 
The lower limit given by the combination of SK-I,II,III and SK-IV turns out to be smaller than that from only SK-IV data. This can be explained due to two candidates in the final $K^0_S\to2\pi^0$ selection sample in SK-II data compared to a background expectation of 0.20 events, which corresponds to a local $p$-value of 1.8\%.
No candidate events were found in the $K^0_S\to2\pi^0$ selection in SK-IV.

\section{Conclusion}
\label{sec:conclusion}
Proton decay into a muon and a neutral kaon was searched for in Super-Kamiokande.
The sensitivity was improved with a new event reconstruction algorithm, neutron tagging and optimized selection criteria for each $K^0_S$ and $K^0_L$ decay mode. 
As a result, no significant event excess has been observed in the final sample of 0.2\,Mton$\cdot$years of data in SK-IV. 
From this result, a lower limit of $4.5\times10^{33}$\,years on the lifetime of $p\to\mu^+K^0$ was obtained at 90\% C.L.
By combining with the previous results, a lower limit of $3.6\times 10^{33}$\,years was obtained from 0.37\,Mton$\cdot$years data collected in SK-I to SK-IV.
This limit is more than twice as long as the previous result and the most stringent to data for this channel.

\section*{Acknowledgements}


We gratefully acknowledge cooperation of the Kamioka Mining and Smelting Company.
The Super‐Kamiokande experiment was built and has been operated with funding from the
Japanese Ministry of Education, Science, Sports and Culture, 
and the U.S. Department of Energy.


We gratefully acknowledge the cooperation of the Kamioka Mining and Smelting Company.
The Super‐Kamiokande experiment has been built and operated from funding by the 
Japanese Ministry of Education, Culture, Sports, Science and Technology, the U.S.
Department of Energy, and the U.S. National Science Foundation. Some of us have been 
supported by funds from the National Research Foundation of Korea NRF‐2009‐0083526
(KNRC) funded by the Ministry of Science, ICT, and Future Planning and the Ministry of
Education (2018R1D1A1B07049158, 2021R1I1A1A01059559), 
the Japan Society for the Promotion of Science, the National
Natural Science Foundation of China under Grants No.11620101004, the Spanish Ministry of Science, 
Universities and Innovation (grant PGC2018-099388-B-I00), the Natural Sciences and 
Engineering Research Council (NSERC) of Canada, the Scinet and Westgrid consortia of
Compute Canada, the National Science Centre (UMO-2018/30/E/ST2/00441) and the Ministry
of Education and Science (DIR/WK/2017/05), Poland,
the Science and Technology Facilities Council (STFC) and GridPPP, UK, the European Union's 
Horizon 2020 Research and Innovation Programme under the Marie Sklodowska-Curie grant
agreement no.754496, H2020-MSCA-RISE-2018 JENNIFER2 grant agreement no.822070, and 
H2020-MSCA-RISE-2019 SK2HK grant agreement no. 872549.

\appendix

\section{$p\to\mu^+K^0$ search in SK-I to SK-III}
Table \ref{tab:results_SK123} shows the summary of $p\to\mu^+K^0$ search in SK-I to SK-III.
These systematic uncertainties are reevaluated due to the updated kaon interaction model.
The selection criteria of SK-I, -II and SK-III \cite{Regis} are different from those of SK-IV.
\begin{table*}[htbp]
    \centering
    \caption{Summary of the $p\to\mu^+K^0$ search in SK-I to SK-III.}
    \vspace{0.2cm}
    \begin{tabular}{ccccc} \hline\hline
      Detector (exposure)&Search mode & Efficiency (\%) & Background (events) & Candidates (events) \\ \hline
      SK-I (91.7 kton·years)& $K^0_S\to2\pi^0$&$7.0\pm0.7$&$0.37\pm0.05$&0 \\
      &$K^0_S\to\pi^+\pi^-$ (2 ring sample)&$10.6\pm1.1$&$3.0\pm0.5$&6\\
      &$K^0_S\to\pi^+\pi^-$ (3 ring sample)&$2.5\pm0.3$&$0.12\pm0.08$&0\\
      &$K^0_L$&$3.8\pm1.8$&$3.5\pm1.1$&2 \\
      \hline
      SK-II (49.2 kton·years) & $K^0_S\to2\pi^0$&$6.2\pm0.8$&$0.20\pm0.05$&2 \\
      &$K^0_S\to\pi^+\pi^-$ (2 ring sample)&$10.3\pm1.3$&$1.6\pm0.4$&0\\
      &$K^0_S\to\pi^+\pi^-$ (3 ring sample)&$2.4\pm0.3$&$0.23\pm0.08$&1\\
      &$K^0_L$&$3.3\pm1.6$&$1.4\pm0.5$&0 \\
      \hline
      SK-III (31.9 kton·years)& $K^0_S\to2\pi^0$&$6.7\pm0.9$&$0.19\pm0.04$&0 \\
      &$K^0_S\to\pi^+\pi^-$ (2 ring sample)&$10.3\pm1.9$&$1.2\pm0.2$&1\\
      &$K^0_S\to\pi^+\pi^-$ (3 ring sample)&$3.0\pm0.3$&$0.09\pm0.02$&0\\
      &$K^0_L$&$3.8\pm1.8$&$1.3\pm0.6$&1 \\
      \hline \hline
    \end{tabular}
  \label{tab:results_SK123}
\end{table*}

\end{document}